\documentclass[onecolumn]{aastex6}

\usepackage{color}

\newcommand{\logg}{$\log\,g$}

\newcommand{\teff}{$T_{\!\mbox{\scriptsize eff}}$}
\newcommand{\teffq}{$T_{\!\mbox{\scriptsize  eff}}^4$}

\slugcomment{}

\shorttitle{NGC\,55}
\shortauthors{Kudritzki et al.}

\begin{document}
\title{A spectroscopic study of blue supergiant stars in the Sculptor galaxy NGC\,55: chemical evolution and distance.}

\author{R.P. Kudritzki\altaffilmark{1}}
\affil{Institute for Astronomy, University of Hawaii, 2680 Woodlawn Drive, Honolulu, HI 96822, USA}

\author{N. Castro}
\affil{Department of Astronomy, University of Michigan, 1085 S. University Avenue, Ann Arbor, MI 48109, USA}

\author{M. A. Urbaneja}
\affil{Institut f\"ur Astro- und Teilchenphysik, Universit\"at Innsbruck,
    Technikerstr. 25/8, 6020 Innsbruck, Austria}

\author{I.-T. Ho}
\affil{Institute for Astronomy, University of Hawaii, 2680 Woodlawn Drive, Honolulu, HI 96822, USA}

\author{F. Bresolin}
\affil{Institute for Astronomy, University of Hawaii, 2680 Woodlawn Drive, Honolulu, HI 96822, USA}

\author{W. Gieren }
\affil{Departamento de Astronom\'{\i}a, Universidad de Concepci\'on, Casilla 160-C, Concepci\'on, Chile} 

\author{G. Pietrzy\'nski\altaffilmark{2}}
\affil{Departamento de Astronom\'{\i}a, Universidad de Concepci\'on, Casilla 160-C, Concepci\'on, Chile} 

\and

\author{N. Przybilla}
\affil{Institut f\"ur Astro- und Teilchenphysik, Universit\"at Innsbruck,
    Technikerstr. 25/8, 6020 Innsbruck, Austria}

\altaffiltext{1}{University Observatory Munich, Scheinerstr. 1, D-81679 Munich, Germany}
\altaffiltext{2}{Nicolaus Copernicus Astronomical Center, Polish Academy of Sciences, ul. Bartycka 18, PL-00-716 Warszawa, Poland}



\begin{abstract}
Low resolution (4.5 to 5 \AA) spectra of 58 blue supergiant stars distributed over the disk of the Magellanic spiral galaxy NGC\,55 in the Sculptor group
are analyzed by means of non-LTE techniques to determine stellar temperatures, gravities and metallicities (from iron peak and $\alpha$-elements). A metallicity gradient of $-0.22 \pm0.06$ dex/R$_{25}$
is detected. The central metallicity on a logarithmic scale relative to the Sun is [Z] = $-0.37 \pm 0.03$. A chemical evolution model using the observed distribution of stellar and interstellar medium gas
mass column densities reproduces the observed metallicity distribution well and reveals a recent history of strong galactic mass accretion and wind outflows with accretion and mass-loss rates
of the order of the star formation rate. There is an indication of spatial inhomogeneity in metallicity. In addition, the relatively high central metallicity of the disk 
confirms that two extra-planar metal poor HII regions detected in previous work 1.13 to 2.22 kpc above the galactic plane are ionized by massive stars formed in-situ outside the disk. For a sub-sample 
of supergiants, for which Hubble Space Telescope photometry is available, the flux-weighted gravity--luminosity realionship is used to determine a distance modulus of $26.85 \pm 0.10$ mag.
\end{abstract}


\section{Introduction}

The galaxies in the Sculptor group seem to form a filament extended along the line of sight \citep{jerjen1998, karachentsev2003} with three distinct subgroups, 
one  around the starburst galaxy NGC\,253 at about 4 Mpc, the other around NGC\,7793 at about the same distance, and the third much closer at half of this distance around NGC\,300.
While NGC\,253, as the largest and most massive galaxy, seems to form the dynamical center of the group, the subgroup around NGC\,300 appears to be attracted by the gravitational field 
of the Local Group.

NGC\,55 is member of the nearby subgroup. The near-IR Cepheid study of the Araucaria collaboration places NGC\,55 and NGC\,300 at roughly the same distance
of 1.9 Mpc \citep{gieren2005b, gieren2008}. These two galaxies have roughly comparable near-IR magnitudes \citep[K$_{tot}$ = 6.25 and 6.38, respectively, see][]{jarrett2003} and mid-IR fluxes
\citep[F$_{3.6 \mu}$ = 2.02 and 1.63 Jy, F$_{4.5\mu}$ = 1.39 and 1.20 Jy, see][]{dale2009} indicating comparable stellar masses. However, while NGC\,300 is a regular spiral galaxy of 
morphological type Scd with a moderate inclination angle (i = 39.9 degrees), NGC\,55 is an almost edge-on (i = 78 degrees) barred spiral of type SB(s)m with the bar apparently oriented
along the line of sight \citep{devaucouleurs1961, devaucouleurs1991}, which closely resembles the Large Magellanic Cloud \citep{westmeier2013, robinson1964}.

NGC\,300 has been subject to many very detailed studies of its stellar populations, the ISM (atomic and molecular gas distribution, HII regions, planetary nebulae, supernovae remnants, 
dust content) and the very extended faint stellar disk (see \citealt{bresolin2009}, \citealt{vlajic2009}, \citealt{westmeier2011}, \citealt{stasinska2013}, \citealt{kang2016}, \citealt{toribio2016}, and references 
therein). In particular, the metallicity of the ISM and the young stellar population has been investigated by detailed quantitative spectroscopic 
studies of blue supergiant stars \citep{kudritzki2008}, red supergiants \citep{gazak2015}, and HII regions \citep{bresolin2009}. While these three investigations used entirely independent methods 
for metallicity diagnostics, the results with respect to central metallicity and metallicity gradient agreed extremely well. On the other hand, only a handful of HII regions in the disk
of NGC\,55 have been studied todate, with an uncertain range in metallicity between [Z] = $-0.6$ to $-0.2$\footnote{We transform the nebular oxygen abundances (O/H) to metallicity relative to solar [Z] adopting 12\,+\,log(O/H)$_\odot$ = 8.69 from \citep{asplund2009}} \citep{webster1983, stasinska1986, zaritsky1994, tuellmann2003, castro2012, pilyugin2014}. 
\citet{tuellmann2003} also investigated two extra-planar HII regions located 0.8 and 1.5 kpc above the disk and found a metallicity almost a factor of ten smaller than solar. 
\citet{castro2008} described 
spectra and spectral morphology of a large sample of hot massive stars, mostly blue supergiants, which were obtained within the Auracaria collaboration \citep[see][]{gieren2005a},
but so far only a small fraction of these objects, 12 supergiants of early-B spectral type, have been subject to a quantitative spectral analysis \citep{castro2012}. This work indicated 
an average metallicity very similar to the LMC, but remained inconclusive about the possibility of a spatial trend in metallicity, in particular a radial metallicity gradient.

NGC\,55, a galaxy with significant star formation, is very likely subject to mass accretion and gas outflows \citep{westmeier2013, tuellmann2003} and, thus, accurate information about metallicity
and a potential metallicity gradient might allow to constrain the rates of matter inflow and outflow \citep[see][]{kudritzki2015}. We have, therefore, resumed the analysis of the spectra obtained
by \citet{castro2008}, this time focussing on the supergiants of spectral type B8 to A5, for which the signal-to-noise ratio was sufficient for a quantitative spectral analysis. Because of the many
metal lines in their spectra and because of their enormous intrinsic brightness, supergiants of these spectral types are ideal for extragalactic metallicity studies (see \citealt{kudritzki2008,kudritzki2012,kudritzki2014}, and references therein). The selection of targets resulted in a sample of 46 objects distributed over a large range of galactocentric distances. These objects were then analyzed
in detail with respect to their effective temperatures, gravities and metallicities and the results are presented in this paper. After a brief description in Section 2 of the observations,
the analysis method and the geometrical model used for a de-projection of the location of the targets in the galactic disk, we summarize the results in Section 3. Section 4 focusses on
the metallicity and the metallicity gradient and applies a chemical evolution model.

Blue supergiants, as the brightest stars in the Universe at optical wavelengths, are also excellent distance indicators, because their 
``flux-weighted gravity'' $g_F\,\equiv\,g$/{\teffq} (\teff\ in units of 10$^{4}$~K) is tightly correlated with their absolute bolometric magnitude, leading to the
``Flux-weighted Gravity -- Luminosity Relationship (FGLR)'' \citep[see][]{kudritzki2003, kudritzki2008}. Since the distance moduli to NGC\,55 obtained with different methods
appear to be slightly controversial ranging from 26.4 mag \citep[Cepheids:][]{gieren2008} to 26.6 mag \citep[EDD database, http;//edd.ifa.hawaii.edu, see][]{tully2009} to 26.8 mag 
\citep[PLNF:][]{vansteene2006}, we select a sub-sample of 13 supergiants, for which HST/ACS photometry is available, and determine an independent distance in Section 5 using the most 
recent calibration of the FGLR method by \citet{urbaneja2016}. Section 6 presents a final discussion.

\section{Observational data and analysis method}
\subsection{Target selection, spectra and photometry}

The spectroscopic observations were carried out in 2004 November 6, 7, 10 using the focal reducer low-dispersion spectrograph FORS2 attached to the ESO VLT-UT2. 
The Mask eXchange Unit (MXU) mode was used for the multi-object spectroscopy with slit widths of 1 arcsecond in conjunction with the 600B grism providing a nominal resolution of 5~\AA~ in 
the wavelength range from 3100 to 6210~\AA. Four fields (A, B, C, D) with a field of view of $6.8 \times 6.8$ arcmin$^2$ were chosen to cover the whole galaxy, and a total of 200 objects were observed. 
For the central field C two different mask settings were applied because of the high density of targets in the center of the galaxy. Target selection, field settings and details of the 
observations, as well as the data reduction, are described in \citet{castro2008}. Their paper provides comprehensive information for all targets. For our work we use coordinates, spectral type, radial
velocity, signal-to-noise ratio and the reduced normalized spectra as displayed in the appendix of Castro et al.. We have applied small corrections to some of the Castro et al.~radial velocities based on the comparison with our model atmosphere spectra and, therefore, give radial velocity values for all our objects. After careful inspection of the digital spectrum of every 
target with spectral type B8 to A5 we
finally selected 46 objects with spectra suitable for a quantitative spectral analysis. Objects with too strong nebular HII emission contaminating the stellar Balmer lines and with spectra too noisy for the 
analysis were dropped from the sample in the selection process. The final list is given in Table~\ref{table_asg}.

Photometry is needed to determine reddening, extinction and apparent bolometric magnitudes which can then be used for a determination of distances through the FGLR method.
As pointed out by \citet{castro2012} the original ground-based photometry obtained with the Warsaw 1.3m telescope at Las Campanas Observatory and provided in the 2008 paper was 
affected by calibration issues. Re-calibrated photometry, which was used for the early B-supergiant targets in \citet{castro2012}, is available for almost all our targets. However, 
a comparison with a sub-sample of 13 stars, for which HST/ACS photometry obtained within the ANGST treasury survey of nearby galaxies \citep{dalcanton2009} is available, revealed large differences 
in both the $V$ and $I$ bands of a few tenths of a magnitude. We, thus, decided to only use the sub-sample of 13 stars for the 
distance determination. The photometry for these objects, together with the information needed for the distance determination, is given given in Table~\ref{table_asg_hst}.

\citet{castro2012} have analyzed 12 early B-supergiants of spectral type B0 to B5 and determined effective temperature, gravity and metallicity. In our investigation of the metallicity and
the metallicity gradient we include the results from their paper and add the metallicities to enlarge the sample. The list of these objects together with their stellar parameters, galactic 
positions, galactocentric distances and radial velocities is provided in Table~\ref{table_bsg}. For one of
these objects, C1 13, HST photometry is available. It is, therefore, included in the sub-sample used for the distance determination.

\subsection{Geometric and kinematic model of the disk of NGC\,55}

The investigation of the metallicity distribution accross the disk of NGC\,55 as a function of galactocentric radius requires a de-projection of the targets into the intrinsic plane of the
galactic disk in order to calculate galactocentric distances. We use the information given in Table~\ref{table_asg} for the de-projection. Our values for the position angle and
inclination were guided by the disk model developed by \citet{puche1991} and are simplified averages of the inner disk. The range of $\pm$ 4 degrees in the inclination angle is used to estimate
the uncertainties in deprojected galactocentric distance for the individual objects. We note that \citet{westmeier2013} have developed an improved model of 
the warped disk but the changes relative to Puche et al. are relatively small in the inner disk, where our targets are located, and are mostly important in the outer disk, for 
which Westmeier et al. present new data. De-projected coordinates y perpendicular to the major axis and x along the major axis are given in Table~\ref{table_asg} and \ref{table_bsg} together
with galactocentric distances in units of the isophotal radius R$_{25}$.

With the de-projected location of our targets we can also compare the radial velocities of our targets. For this purpose we use the kinematic models and rotation curves by \citet{puche1991} and 
\citet{westmeier2013} together with the systemic velocity of Table~\ref{table_asg}. We note that\citet{westmeier2013} have found that the systemic velocity given by \citet{puche1991} is in error and 
we use their systemic velocity.

\subsection{Spectral analysis method}

The analysis method is described in detail in \citet{kudritzki2013, kudritzki2014} and in \citet{hosek2014}. In short, we compare normalized observed spectra with synthetic spectra drawn from a 
comprehensive grid of metal line-blanketed model atmospheres with extensive non-LTE line formation calculations using elaborate atomic models (\citealt{przybilla2006}). The grid of models is described in 
\citet{kudritzki2008, kudritzki2012}. The fit procedure uses the higher Balmer lines (usually H$_{5,6,8,9,10}$) in a first step to constrain gravity as a function of effective temperature. 
This yields a relationship between gravity and effective temperature for each star (see Fig.~3 in \citealt{kudritzki2014}). Along this relationship gravity is usually constrained
with an accuracy of 0.05 dex. 

Then, 
in a second step, up to 11 spectral windows in the range from 3990 to 6000 \AA~dominated by metal lines are selected and a comparison between observed and calculated flux is carried out
as a function of metallicity for each point along the gravity-temperature relationship (metallicity [Z] is defined as [Z] = log Z/Z$_{\odot}$, where Z$_{\odot}$ is the solar metallicity). 
In this way, the quality of the metal line fits can be assessed by calculating a $\chi^2$ value
for each temperature and metallicity. The minimum of $\chi^2$ and the $\Delta \chi^2$-isocontours around the minimum then yield metallicity and effective temperature of the star with the corresponding
fit uncertainties. The final gravity corresponds to the gravity of the gravity-temperature relationship at the temperature determined from the $\chi^2$ minimum. The fit uncertainties of the gravities
are a combination of the 0.05 dex accuracy at each temperature as mentioned above and the uncertainty in temperature. The next section will show a few examples of the spectral fits obtained.
Note that the $\chi^2$-fit determines stellar metallicity from the contribution of many elements, including the iron peak and $\alpha$-elements.

In the fit procedure we take into account the fact that the actual spectral resolution can be better than the nominal resolution of the spectrograph in situations of good atmospheric seeing. We, thus, measured the 
effective resolution
for the set of spectra in each field. We found that the spectra of fields A, B, and C are well described with a resolution of 4.5~\AA, whereas the resolution obtained for field D is 5~\AA. The
synthetic spectra were degraded to these resolution values accordingly.

\clearpage
\begin{deluxetable}{ c c c c c c c c c c }
\tablecaption{NGC\,55 late B and early A supergiants \label{table_asg}}
\tablehead{
\colhead{Star} & sp.  & \colhead{\teff} & \colhead{\logg} & \colhead{\logg$_{F}$} &  \colhead{[Z]} & \colhead{R/R$_{25}$} & \colhead{x/R$_{25}$} & \colhead{y/R$_{25}$} &  \colhead{v$_{rad}$ } \\
\colhead{ }    & type & \colhead{K}     & \colhead{cgs}   & \colhead{cgs}        &  \colhead{dex}  & \colhead{ }         &  \colhead{ }       & \colhead{ }          & \colhead{km/s}       \\[1mm]
\colhead{(1)}  & \colhead{(2)}	 & \colhead{(3)}   & \colhead{(4)}	 &  \colhead{(5)} & \colhead{(6)}         &  \colhead{(7)}     & \colhead{(8)}          & \colhead{(9)} & \colhead{(10)}
}
\startdata
 D 29  & A3 I &  8750$\rm^{+150}_{-200}$ & 1.15 & 1.38$\rm^{+0.09}_{-0.09}$ & $-0.50\rm^{+0.10}_{-0.20}$ & 0.290$\rm^{+0.000}_{-0.000}$ & $-0.290$ & $-0.007\rm^{+0.002}_{-0.003}$ &  98. \\
 D 35  & A3 I &  8400$\rm^{+150}_{-200}$ & 1.00 & 1.30$\rm^{+0.10}_{-0.12}$ & $-0.20\rm^{+0.15}_{-0.10}$ & 0.299$\rm^{+0.062}_{-0.025}$ & $-0.237$ & $-0.182\rm^{+0.045}_{-0.090}$ &  63. \\
 D 39  & A2 I &  9900$\rm^{+200}_{-250}$ & 1.23 & 1.25$\rm^{+0.05}_{-0.06}$ & $-0.40\rm^{+0.08}_{-0.08}$ & 0.200$\rm^{+0.009}_{-0.003}$ & $-0.193$ & +0.053$\rm^{+0.026}_{-0.013}$ & 132. \\
 C1 1  & A0 I &  9450$\rm^{+250}_{-250}$ & 1.34 & 1.43$\rm^{+0.07}_{-0.08}$ & $-0.45\rm^{+0.10}_{-0.10}$ & 0.200$\rm^{+0.038}_{-0.015}$ & $-0.163$ & +0.116$\rm^{+0.057}_{-0.029}$ & 144. \\ 
 C1 10 & A2 I &  8700$\rm^{+250}_{-250}$ & 1.03 & 1.27$\rm^{+0.10}_{-0.09}$ & $-0.45\rm^{+0.07}_{-0.13}$ & 0.171$\rm^{+0.073}_{-0.034}$ & $-0.069$ & +0.157$\rm^{+0.077}_{-0.039}$ & 100. \\
 C1 17 & A3 I &  8500$\rm^{+70}_{-70}$   & 1.15 & 1.43$\rm^{+0.08}_{-0.07}$ & $-0.55\rm^{+0.05}_{-0.08}$ & 0.258$\rm^{+0.127}_{-0.064}$ & +0.001 & +0.258$\rm^{+0.127}_{-0.063}$ & 117. \\
 C2 15 & A0 I & 10500$\rm^{+250}_{-200}$ & 1.60 & 1.59$\rm^{+0.06}_{-0.06}$ & $-0.42\rm^{+0.08}_{-0.08}$ & 0.194$\rm^{+0.094}_{-0.046}$ & $-0.031$ & +0.192$\rm^{+0.095}_{-0.047}$ &  98. \\
 C2 13 & A2 I &  9000$\rm^{+200}_{-200}$ & 1.25 & 1.43$\rm^{+0.09}_{-0.09}$ & $-0.37\rm^{+0.05}_{-0.10}$ & 0.096$\rm^{+0.034}_{-0.016}$ & $-0.054$ & $-0.079\rm^{+0.019}_{-0.039}$ &  90. \\
 C2 39 & A2 I &  9000$\rm^{+250}_{-100}$ & 1.20 & 1.38$\rm^{+0.09}_{-0.07}$ & $-0.60\rm^{+0.08}_{-0.08}$ & 0.143$\rm^{+0.034}_{-0.015}$ & +0.107 & $-0.094\rm^{+0.023}_{-0.047}$ & 158. \\
 B 7   & A0 I &  9500$\rm^{+100}_{-100}$ & 1.15 & 1.24$\rm^{+0.06}_{-0.06}$ & $-0.40\rm^{+0.05}_{-0.05}$ & 0.260$\rm^{+0.026}_{-0.010}$ & +0.237 & $-0.108\rm^{+0.026}_{-0.053}$ & 162. \\
 A 38  & A0 I &  9650$\rm^{+250}_{-250}$ & 1.33 & 1.39$\rm^{+0.06}_{-0.07}$ & $-0.52\rm^{+0.08}_{-0.08}$ & 0.896$\rm^{+0.005}_{-0.001}$ & +0.893 & $-0.081\rm^{+0.020}_{-0.040}$ & 147. \\
 A 10  & A0 I &  9550$\rm^{+100}_{-100}$ & 1.41 & 1.49$\rm^{+0.06}_{-0.06}$ & $-0.67\rm^{+0.05}_{-0.05}$ & 0.735$\rm^{+0.093}_{-0.035}$ & +0.650 & +0.343$\rm^{+0.169}_{-0.084}$ & 187. \\
 A 43  & A2 I &  9000$\rm^{+250}_{-250}$ & 1.15 & 1.33$\rm^{+0.09}_{-0.08}$ & $-0.60\rm^{+0.10}_{-0.10}$ & 0.969$\rm^{+0.011}_{-0.004}$ & +0.960 & $-0.132\rm^{+0.033}_{-0.065}$ & 185. \\
 A 6   & A3 I &  8650$\rm^{+150}_{-150}$ & 1.18 & 1.43$\rm^{+0.10}_{-0.10}$ & $-0.25\rm^{+0.10}_{-0.10}$ & 0.613$\rm^{+0.006}_{-0.002}$ & +0.608 & $-0.079\rm^{+0.019}_{-0.039}$ & 288. \\
 A 5   & A3 I &  8750$\rm^{+50}_{-50}$   & 1.45 & 1.68$\rm^{+0.07}_{-0.07}$ & $-0.30\rm^{+0.05}_{-0.05}$ & 0.600$\rm^{+0.004}_{-0.001}$ & +0.597 & +0.061$\rm^{+0.030}_{-0.015}$ & 260. \\
 A 28  & A2 I &  9150$\rm^{+150}_{-150}$ & 1.67 & 1.83$\rm^{+0.09}_{-0.09}$ & $-0.30\rm^{+0.13}_{-0.13}$ & 0.789$\rm^{+0.004}_{-0.002}$ & +0.785 & +0.061$\rm^{+0.072}_{-0.036}$ & 219. \\
 A 33  & A0 I &  9800$\rm^{+200}_{-100}$ & 1.86 & 1.90$\rm^{+0.07}_{-0.06}$ & $-0.54\rm^{+0.10}_{-0.10}$ & 0.881$\rm^{+0.047}_{-0.018}$ & +0.840 & +0.265$\rm^{+0.131}_{-0.065}$ & 199. \\
 A 25  & B8 I & 12200$\rm^{+250}_{-250}$ & 1.78 & 1.43$\rm^{+0.05}_{-0.05}$ & $-0.57\rm^{+0.10}_{-0.10}$ & 0.923$\rm^{+0.171}_{-0.067}$ & +0.757 & +0.529$\rm^{+0.261}_{-0.130}$ & 229. \\
 A 14  & B9 I & 10350$\rm^{+150}_{-150}$ & 1.37 & 1.31$\rm^{+0.05}_{-0.05}$ & $-0.45\rm^{+0.10}_{-0.10}$ & 0.701$\rm^{+0.000}_{-0.001}$ & +0.700 & +0.027$\rm^{+0.013}_{-0.007}$ & 225. \\
 A 9   & A5 I &  7900$\rm^{+50}_{-50}$   & 0.85 & 1.26$\rm^{+0.09}_{-0.09}$ & $-0.65\rm^{+0.07}_{-0.07}$ & 0.658$\rm^{+0.018}_{-0.007}$ & +0.643 & $-0.141\rm^{+0.035}_{-0.069}$ & 228. \\
 A 4   & A5 I &  8200$\rm^{+150}_{-150}$ & 1.08 & 1.42$\rm^{+0.13}_{-0.13}$ & $-0.30\rm^{+0.10}_{-0.10}$ & 0.586$\rm^{+0.001}_{-0.001}$ & +0.585 & +0.035$\rm^{+0.017}_{-0.009}$ & 256. \\
 C1 14 & A5 I &  8200$\rm^{+100}_{-100}$ & 0.97 & 1.32$\rm^{+0.10}_{-0.10}$ & $-0.45\rm^{+0.10}_{-0.10}$ & 0.187$\rm^{+0.090}_{-0.044}$ & $-0.031$ & $-0.184\rm^{+0.045}_{-0.045}$ & 107. \\
 C1 52 & A2 I &  8400$\rm^{+150}_{-150}$ & 0.80 & 1.10$\rm^{+0.08}_{-0.08}$ & $-0.41\rm^{+0.09}_{-0.06}$ & 0.263$\rm^{+0.000}_{-0.000}$ & +0.263 & $-0.007\rm^{+0.002}_{-0.004}$ & 149. \\
 C2 54 & B9 I & 11000$\rm^{+300}_{-300}$ & 1.65 & 1.48$\rm^{+0.06}_{-0.05}$ & $-0.23\rm^{+0.13}_{-0.12}$ & 0.275$\rm^{+0.014}_{-0.005}$ & +0.263 & +0.081$\rm^{+0.040}_{-0.020}$ & 223. \\
 C1 49 & A5 I &  8150$\rm^{+200}_{-200}$ & 0.99 & 1.35$\rm^{+0.15}_{-0.17}$ & $-0.45\rm^{+0.13}_{-0.13}$ & 0.242$\rm^{+0.001}_{-0.001}$ & +0.241 & $-0.022\rm^{+0.005}_{-0.011}$ & 165. \\
 C2 51 & B9 I & 10600$\rm^{+300}_{-300}$ & 1.47 & 1.37$\rm^{+0.05}_{-0.05}$ & $-0.38\rm^{+0.10}_{-0.10}$ & 0.237$\rm^{+0.003}_{-0.002}$ & +0.234 & $-0.035\rm^{+0.009}_{-0.017}$ & 213. \\
 C1 47 & A0 I & 10250$\rm^{+250}_{-250}$ & 1.25 & 1.21$\rm^{+0.05}_{-0.05}$ & $-0.28\rm^{+0.10}_{-0.10}$ & 0.232$\rm^{+0.004}_{-0.001}$ & +0.229 & +0.037$\rm^{+0.018}_{-0.009}$ & 182. \\
 C1 42 & A2 I &  9450$\rm^{+250}_{-250}$ & 1.09 & 1.19$\rm^{+0.07}_{-0.07}$ & $-0.28\rm^{+0.10}_{-0.10}$ & 0.208$\rm^{+0.015}_{-0.005}$ & +0.195 & +0.073$\rm^{+0.036}_{-0.018}$ & 149. \\
 C2 44 & A2 I &  8850$\rm^{+200}_{-150}$ & 1.28 & 1.55$\rm^{+0.13}_{-0.11}$ & $-0.50\rm^{+0.10}_{-0.10}$ & 0.382$\rm^{+0.163}_{-0.077}$ & +0.153 & +0.350$\rm^{+0.173}_{-0.086}$ & 142. \\
 C1 34 & B8 I & 11500$\rm^{+300}_{-300}$ & 1.60 & 1.36$\rm^{+0.05}_{-0.05}$ & $-0.40\rm^{+0.10}_{-0.10}$ & 0.116$\rm^{+0.005}_{-0.002}$ & +0.111 & +0.033$\rm^{+0.016}_{-0.008}$ & 176. \\
 C2 37 & A0 I &  9750$\rm^{+350}_{-350}$ & 1.50 & 1.54$\rm^{+0.07}_{-0.08}$ & $-0.36\rm^{+0.12}_{-0.12}$ & 0.092$\rm^{+0.001}_{-0.000}$ & +0.092 & +0.010$\rm^{+0.005}_{-0.002}$ & 168. \\
 C1 32 & B9 I & 10200$\rm^{+200}_{-200}$ & 1.24 & 1.20$\rm^{+0.05}_{-0.05}$ & $-0.66\rm^{+0.05}_{-0.05}$ & 0.501$\rm^{+0.240}_{-0.118}$ & +0.093 & +0.492$\rm^{+0.243}_{-0.121}$ & 171. \\
 C2 29 & B8 I & 12500$\rm^{+500}_{-300}$ & 1.55 & 1.16$\rm^{+0.05}_{-0.05}$ & $-0.26\rm^{+0.12}_{-0.12}$ & 0.038$\rm^{+0.005}_{-0.002}$ & +0.033 & $-0.018\rm^{+0.004}_{-0.009}$ & 160. \\
 C2 22 & A2 I & 10500$\rm^{+300}_{-500}$ & 1.25 & 1.17$\rm^{+0.05}_{-0.05}$ & $-0.30\rm^{+0.20}_{-0.20}$ & 0.175$\rm^{+0.085}_{-0.042}$ & +0.027 & +0.173$\rm^{+0.086}_{-0.043}$ & 140. \\
 C2 19 & A3 I &  8750$\rm^{+200}_{-200}$ & 1.38 & 1.61$\rm^{+0.12}_{-0.12}$ & $-0.35\rm^{+0.07}_{-0.07}$ & 0.067$\rm^{+0.033}_{-0.016}$ & $-0.006$ & +0.067$\rm^{+0.033}_{-0.016}$ &  91. \\
 C1 16 & A5 I &  8200$\rm^{+120}_{-120}$ & 1.03 & 1.37$\rm^{+0.11}_{-0.11}$ & $-0.55\rm^{+0.15}_{-0.15}$ & 0.328$\rm^{+0.162}_{-0.080}$ & $-0.013$ & +0.328$\rm^{+0.162}_{-0.081}$ & 164. \\
 C1 15 & B8 I & 11200$\rm^{+300}_{-300}$ & 1.68 & 1.48$\rm^{+0.05}_{-0.05}$ & $-0.50\rm^{+0.08}_{-0.08}$ & 0.287$\rm^{+0.141}_{-0.070}$ & $-0.024$ & +0.286$\rm^{+0.141}_{-0.070}$ & 158. \\
 C1 11 & A3 I &  8625$\rm^{+175}_{-175}$ & 1.19 & 1.45$\rm^{+0.11}_{-0.10}$ & $-0.48\rm^{+0.10}_{-0.10}$ & 0.132$\rm^{+0.053}_{-0.024}$ & $-0.062$ & $-0.117\rm^{+0.029}_{-0.058}$ & 123. \\
 C2 14 & B8 I & 12000$\rm^{+200}_{-200}$ & 1.65 & 1.33$\rm^{+0.05}_{-0.05}$ & $-0.36\rm^{+0.06}_{-0.06}$ & 0.379$\rm^{+0.185}_{-0.091}$ & $-0.049$ & +0.376$\rm^{+0.186}_{-0.092}$ & 130. \\
 C1 8  & B8 I & 11500$\rm^{+200}_{-300}$ & 1.60 & 1.36$\rm^{+0.05}_{-0.05}$ & $-0.51\rm^{+0.08}_{-0.08}$ & 0.193$\rm^{+0.076}_{-0.035}$ & $-0.093$ & +0.169$\rm^{+0.083}_{-0.042}$ & 149. \\
 C2 8  & A0 I &  9650$\rm^{+150}_{-150}$ & 1.70 & 1.76$\rm^{+0.06}_{-0.07}$ & $-0.66\rm^{+0.05}_{-0.05}$ & 0.392$\rm^{+0.183}_{-0.089}$ & $-0.101$ & +0.379$\rm^{+0.187}_{-0.093}$ &  95. \\
 D 45  & A2 I &  9500$\rm^{+500}_{-500}$ & 1.58 & 1.66$\rm^{+0.09}_{-0.13}$ & $-0.60\rm^{+0.10}_{-0.10}$ & 0.154$\rm^{+0.012}_{-0.005}$ & $-0.142$ & +0.058$\rm^{+0.029}_{-0.014}$ & 129. \\
 C2 1  & A0 I &  9900$\rm^{+250}_{-250}$ & 1.58 & 1.60$\rm^{+0.06}_{-0.07}$ & $-0.55\rm^{+0.10}_{-0.10}$ & 0.163$\rm^{+0.002}_{-0.000}$ & $-0.162$ & $-0.023\rm^{+0.006}_{-0.011}$ &  86. \\
 C2 2  & A2 II &  8850$\rm^{+150}_{-150}$ & 1.71 & 1.93$\rm^{+0.11}_{-0.11}$ & $-0.61\rm^{+0.12}_{-0.12}$ & 0.445$\rm^{+0.196}_{-0.095}$ & $-0.155$ & +0.417$\rm^{+0.206}_{-0.102}$ &  70. \\
 D 36  & B9 I & 10850$\rm^{+250}_{-250}$ & 1.65 & 1.51$\rm^{+0.05}_{-0.05}$ & $-0.40\rm^{+0.08}_{-0.08}$ & 0.313$\rm^{+0.083}_{-0.035}$ & $-0.224$ & +0.218$\rm^{+0.108}_{-0.054}$ &  78. \\
 D 6   & A3 II &  8400$\rm^{+150}_{-150}$ & 1.58 & 1.88$\rm^{+0.13}_{-0.13}$ & $-0.60\rm^{+0.15}_{-0.15}$ & 0.512$\rm^{+0.014}_{-0.005}$ & $-0.501$ & $-0.109\rm^{+0.027}_{-0.054}$ &  58. \\
\enddata
\tablecomments{
NGC\,55 galaxy parameters: R$_{25}$ = 16.18 arcmin \citep{devaucouleurs1991}, position angle PA = 109\arcdeg (see section 2.2), inclination i =78\arcdeg $\pm$ 4\arcdeg (see section 2.2),
central coordinates $\alpha_{2000}$ = 00$^{\rm h}$ 14$^{\rm m}$ 54.$^{\rm s}$009, $\delta_{2000}$ = -39\arcdeg 11\arcmin 49.\arcsec267 \citep{hummel1986,puche1991}.\\
x is coordinate along the major axis, y is the de-projected coordinate along the minor axis.\\
$g_F=g$/{\teffq} (\teff\ in units of 10$^{4}$~K).
}
\end{deluxetable}

\clearpage
\begin{deluxetable}{ c c c c c c c c c c }
\tablecaption{NGC\,55 late O, early and mid B supergiants \label{table_bsg}}
\tablehead{
\colhead{Star} & sp.  & \colhead{\teff} & \colhead{\logg} & \colhead{\logg$_{F}$} &  \colhead{[Z]} & \colhead{R/R$_{25}$} & \colhead{x/R$_{25}$} & \colhead{y/R$_{25}$} &  \colhead{v$_{rad}$ } \\
\colhead{ }    & type & \colhead{K}     & \colhead{cgs}   & \colhead{cgs}        &  \colhead{dex}  & \colhead{ }         &  \colhead{ }       & \colhead{ }          & \colhead{km/s}       \\[1mm]
\colhead{(1)}  & \colhead{(2)}	 & \colhead{(3)}   & \colhead{(4)}	  &  \colhead{(5)}  & \colhead{(6)}       &  \colhead{(7)}     & \colhead{(8)}         & \colhead{(9)} & \colhead{(10)}
}
\startdata
 A  08 & O9.7 I & 27700$\pm$1200 & 2.91 & 1.14$\pm$0.07 & $-0.37\pm$0.15 &  0.634$\rm^{+0.003}_{-0.001}$ & +0.631 & $-0.056\rm^{+0.014}_{-0.028}$ & 304. \\
 C1 44 &  B0 I  & 26200$\pm$1500 & 2.83 & 1.16$\pm$0.07 & $-0.53\pm$0.15 &  0.239$\rm^{+0.030}_{-0.012}$ & +0.211 & $-0.112\rm^{+0.027}_{-0.055}$ & 181. \\
 C1 9  &  B1 I  & 17400$\pm$1500 & 2.37 & 1.41$\pm$0.07 & $-0.51\pm$0.15 &  0.398$\rm^{+0.190}_{-0.093}$ & $-0.081$ & +0.390$\rm^{+0.192}_{-0.096}$ & 118. \\
 C1 13 &  B1 I  & 24200$\pm$1400 & 2.69 & 1.15$\pm$0.09 & $-0.45\pm$0.15 &  0.207$\rm^{+0.099}_{-0.049}$ & $-0.042$ & $-0.203\rm^{+0.050}_{-0.100}$ & 115. \\
 C1 45 &  B1 I  & 21500$\pm$2500 & 2.73 & 1.40$\pm$0.12 & $-0.44\pm$0.15 &  0.220$\rm^{+0.004}_{-0.001}$ & +0.216 & $-0.038\rm^{+0.009}_{-0.019}$ & 179. \\
 A 17  &  B1 I  & 22500$\pm$2300 & 2.85 & 1.44$\pm$0.08 & -0.45$\pm$0.15 &  0.738$\rm^{+0.006}_{-0.002}$ & +0.733 & +0.087$\rm^{+0.043}_{-0.021}$ & 207. \\
 D 27  &  B2 I  & 19400$\pm$1600 & 2.41 & 1.26$\pm$0.10 & $-0.42\pm$0.15 &  0.320$\rm^{+0.004}_{-0.001}$ & $-0.317$ & +0.045$\rm^{+0.022}_{-0.011}$ & 152. \\
 A 27  &  B2 I  & 16300$\pm$1100 & 2.13 & 1.28$\pm$0.09 & $-0.57\pm$0.15 &  0.903$\rm^{+0.134}_{-0.052}$ & +0.778 & $-0.459\rm^{+0.113}_{-0.227}$ & 199. \\
 C1 53 & B2.5 I & 18200$\pm$2000 & 2.54 & 1.50$\pm$0.11 & $-0.21\pm$0.15 &  0.292$\rm^{+0.026}_{-0.010}$ & +0.269 & +0.114$\rm^{+0.056}_{-0.028}$ & 164. \\
 B 31  & B2.5 I & 16700$\pm$900  & 2.09 & 1.20$\pm$0.08 & $-0.28\pm$0.15 &  0.497$\rm^{+0.009}_{-0.003}$ & +0.490 & +0.084$\rm^{+0.042}_{-0.021}$ & 190. \\
 A 26  & B2.5 I & 16100$\pm$1000 & 2.15 & 1.32$\pm$0.09 & $-0.15\pm$0.15 &  0.778$\rm^{+0.010}_{-0.004}$ & +0.770 & $-0.115\rm^{+0.028}_{-0.057}$ & 220. \\
 A 11  &  B5 I  & 14400$\pm$1000 & 1.94 & 1.31$\pm$0.08 & $-0.37\pm$0.15 &  0.691$\rm^{+0.034}_{-0.012}$ & +0.662 & +0.197$\rm^{+0.097}_{-0.048}$ & 209. \\
\enddata
\tablecomments{Data from \citet{castro2012}.\\
NGC\,55 galaxy parameters: R$_{25}$ = 16.18 arcmin \citep{devaucouleurs1991}, position angle PA = 109\arcdeg (see section 2.2), inclination i =78\arcdeg $\pm$ 4\arcdeg (see section 2.2),
central coordinates $\alpha_{2000}$ = 00$^{\rm h}$ 14$^{\rm m}$ 54.$^{\rm s}$009, $\delta_{2000}$ = -39\arcdeg 11\arcmin 49.\arcsec267 \citep{hummel1986,puche1991}.\\
x is coordinate along the major axis, y is the de-projected coordinate along the minor axis.\\
$g_F=g$/{\teffq} (\teff\ in units of 10$^{4}$~K).
}
\end{deluxetable}
\clearpage

\section{Results}

\begin{figure}
\begin{center}
\includegraphics[scale=0.45,angle=90]{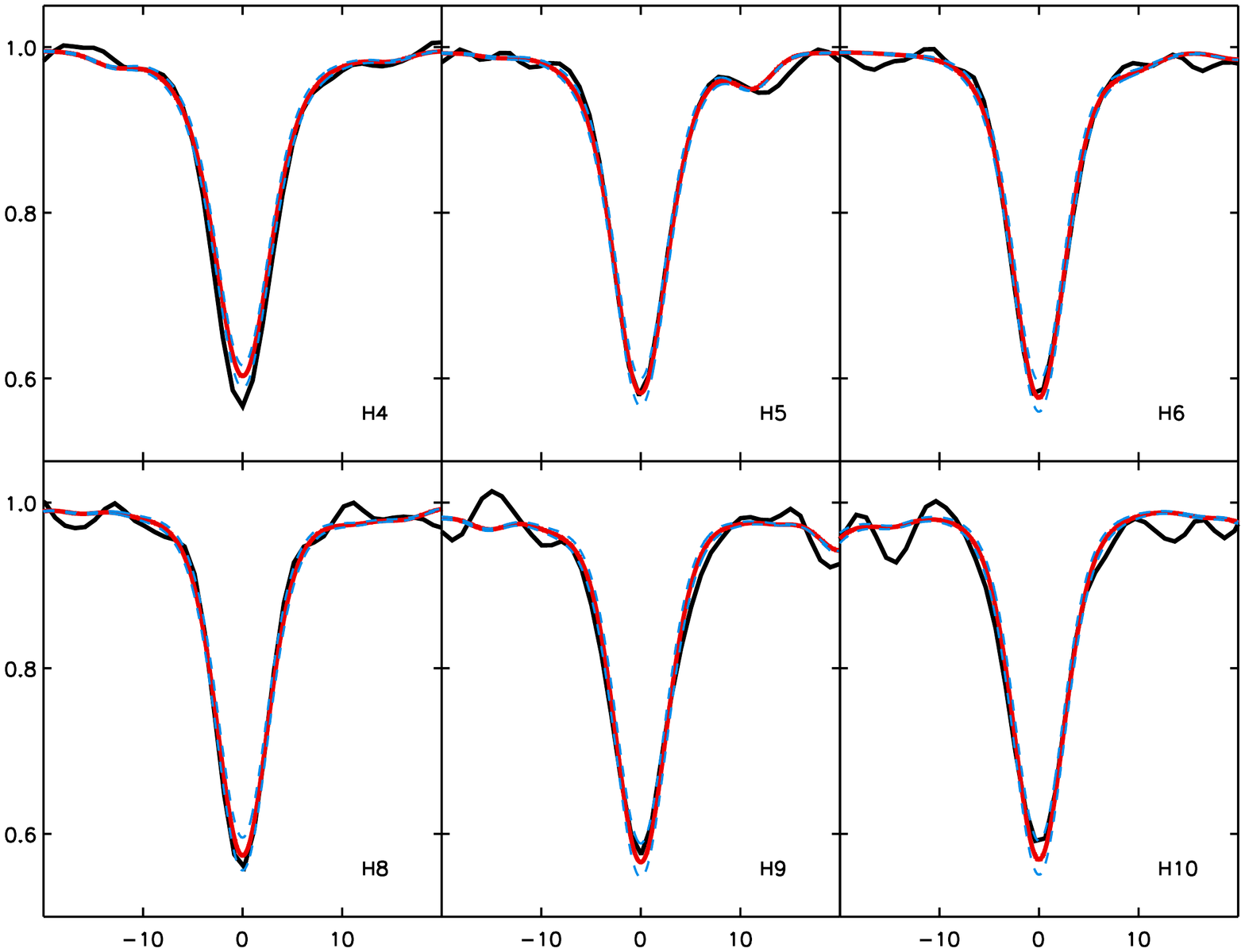}
\includegraphics[scale=0.45,angle=90]{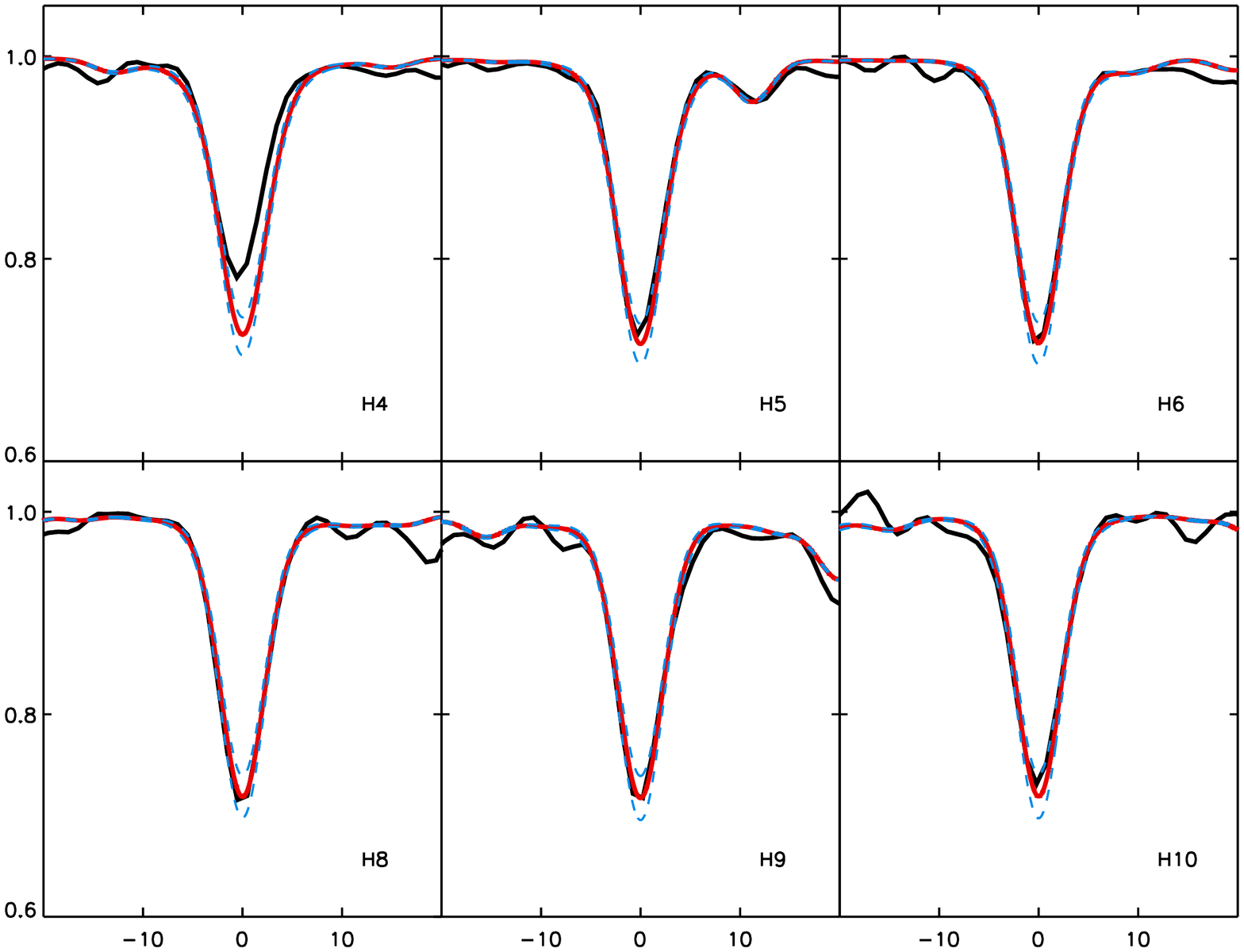}
\caption{\label{balmerfit} Example model fits  of the observed Balmer line profiles for two NGC\,55 supergiant stars (black solid). Synthetic profiles of the final model (red) and two models with gravities 
increased and decreased by 0.05 dex (blue dashed) are overplotted. Top: star C1 49; bottom: star A38.}
 \end{center}
\end{figure}

\begin{figure}
\begin{center}
\includegraphics[scale=0.70]{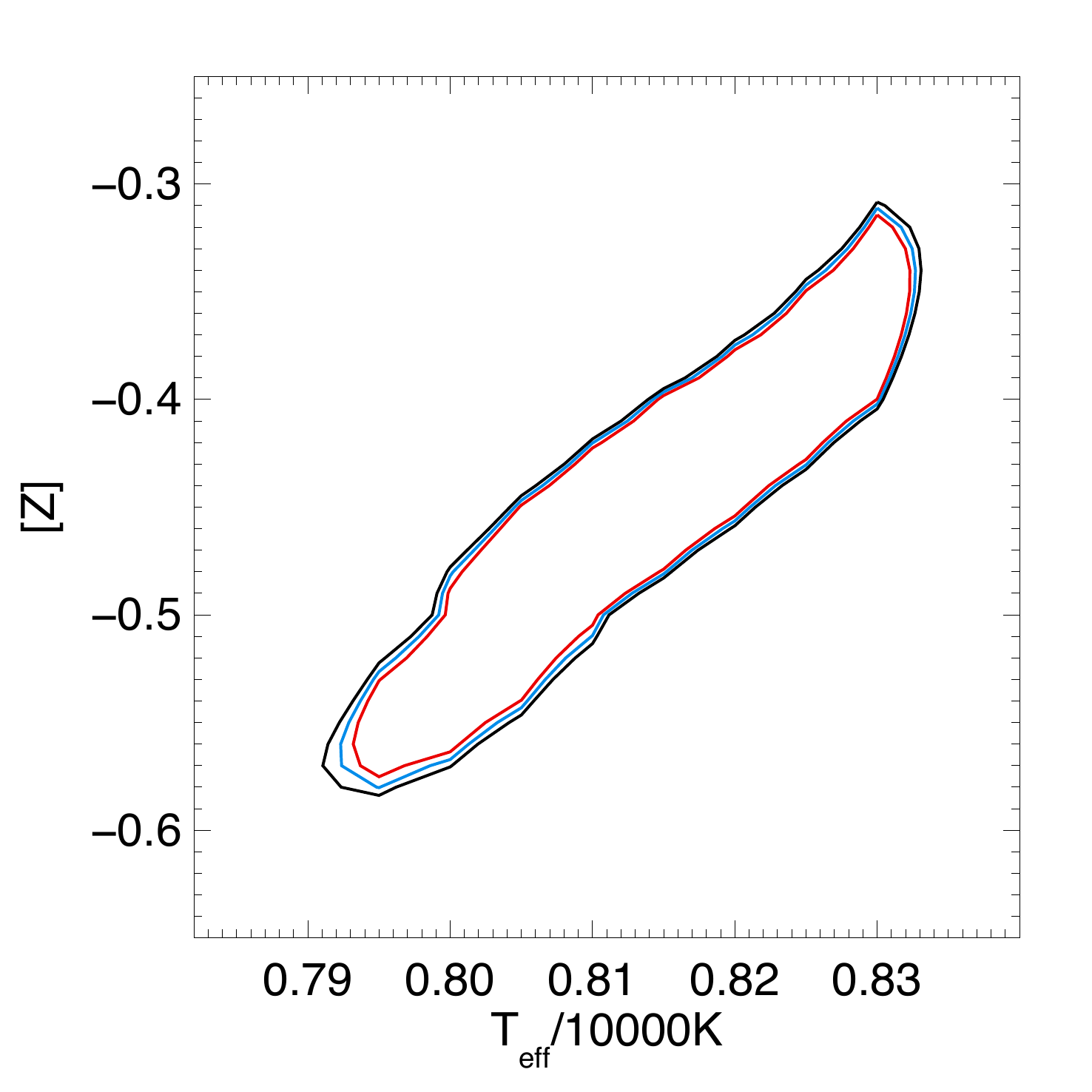}
\includegraphics[scale=0.70]{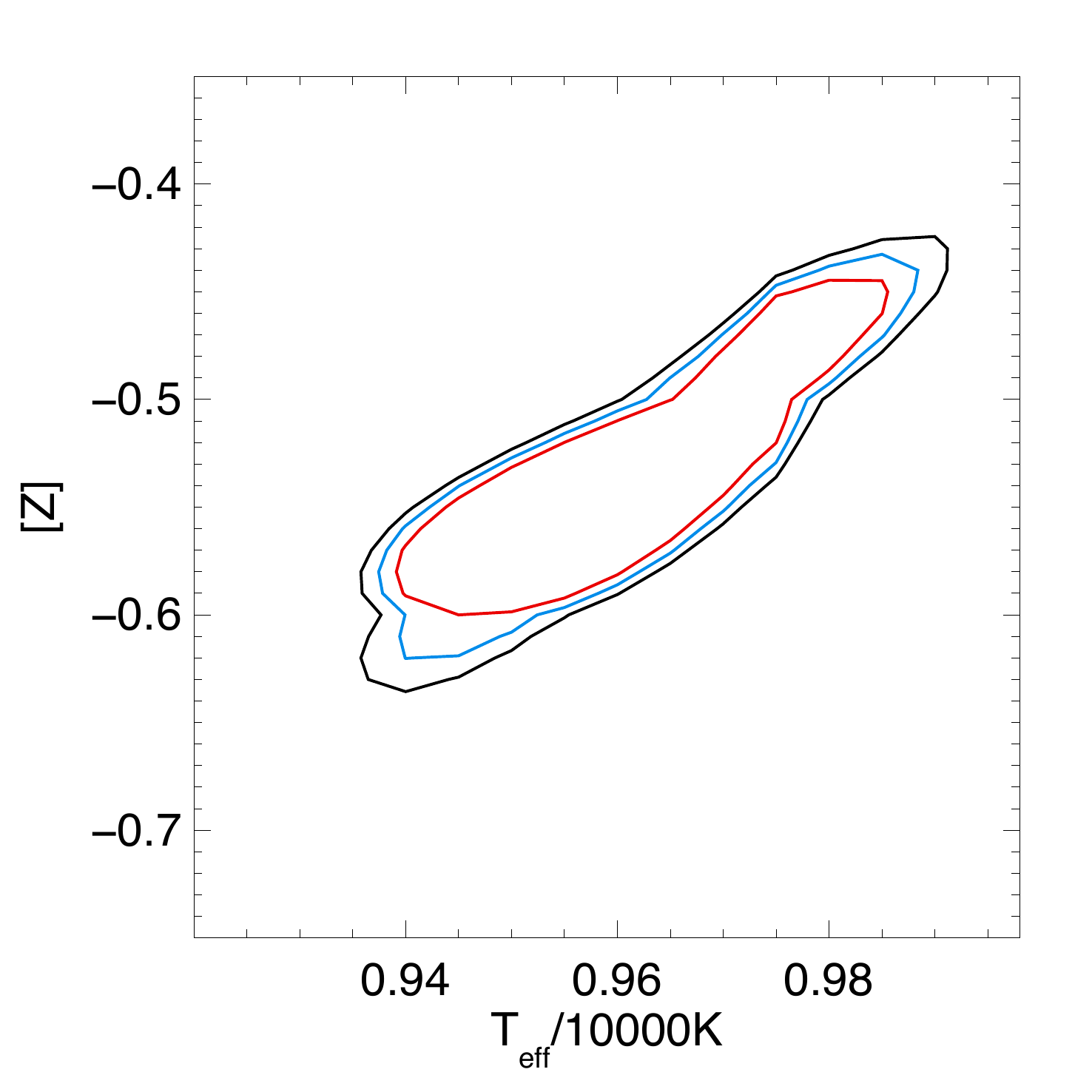}
\caption{\label{chisq} Examples for the determination of effective temperature \teff\ and metallicity [Z] using isocontours, $\Delta \chi^2$, in the metallicity-temperature plane obtained from the
comparison of synthetic with observed spectra (see text). $\Delta \chi^2$ = 3 (red), 6 (blue), and 9 (black), respectively, are plotted. Top: star C1 49; bottom: star A 38. 
}
 \end{center}
\end{figure}

The results of the spectral analysis are given in Table~\ref{table_asg}. Fig.~\ref{balmerfit} shows a fit of the Balmer lines for two stars taken as examples. As for almost all stars in the sample,
H$_{\beta}$ is affected by stellar winds and HII region emission. The other Balmer lines, however, allow for an accurate determination of stellar gravities. Fig.~\ref{chisq} displays
the isocontours of the $\chi^2$ minimization in the metallicity-temperature plane. We see that while there is some degeneracy between metallicity and temperature, both are well constrained. The 
$\Delta \chi^2 = 3$ isocontour corresponds to a 1-$\sigma$ uncertainty, as we have again verified by detailed Monte-Carlo experiments as described by \citet{hosek2014}. This isocontour is used for
all stars to determine the errors for temperature and metallicity in Table~\ref{table_asg}.

\begin{figure*}
\includegraphics[scale=0.50,angle=90]{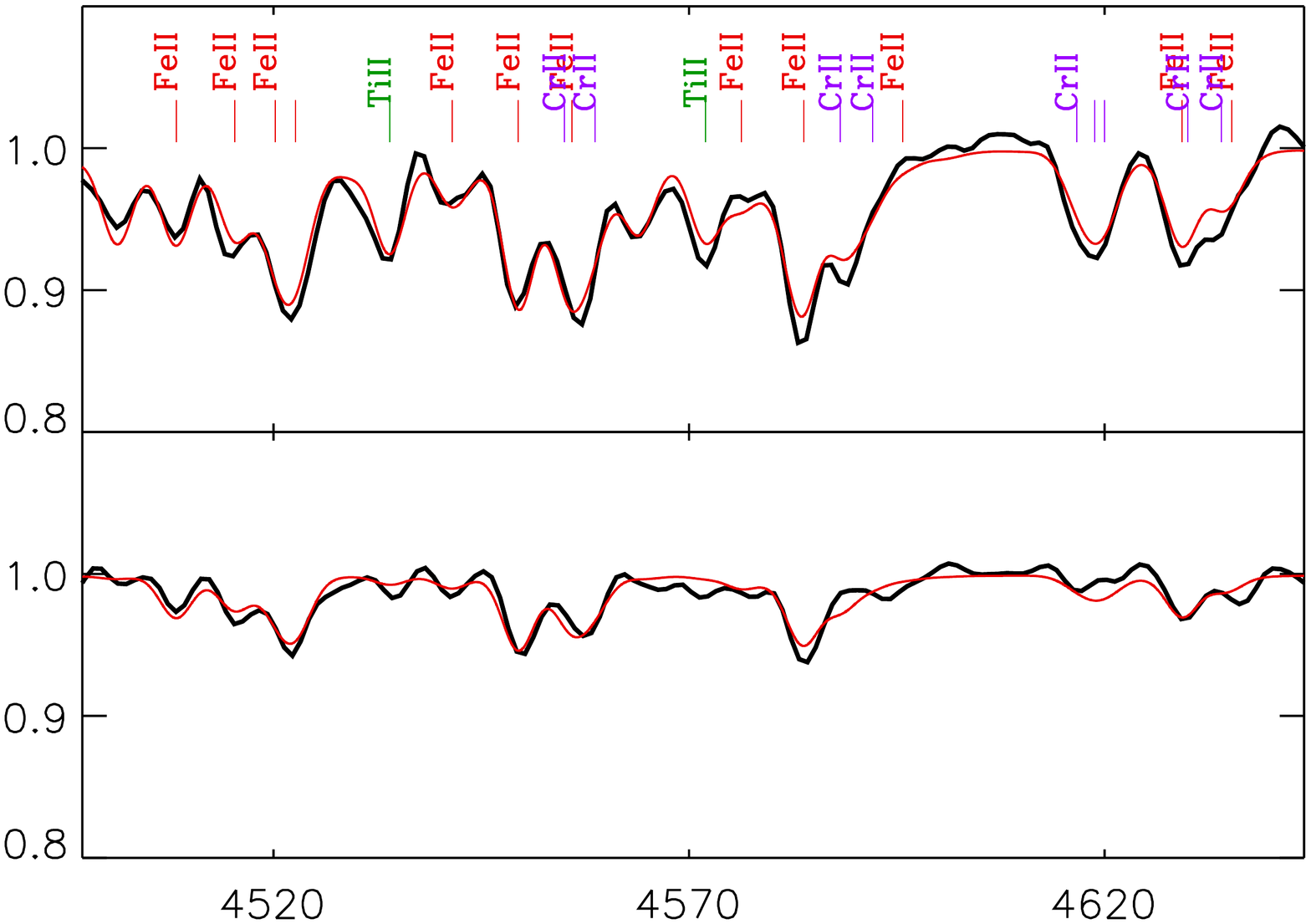}
\includegraphics[scale=0.50,angle=90]{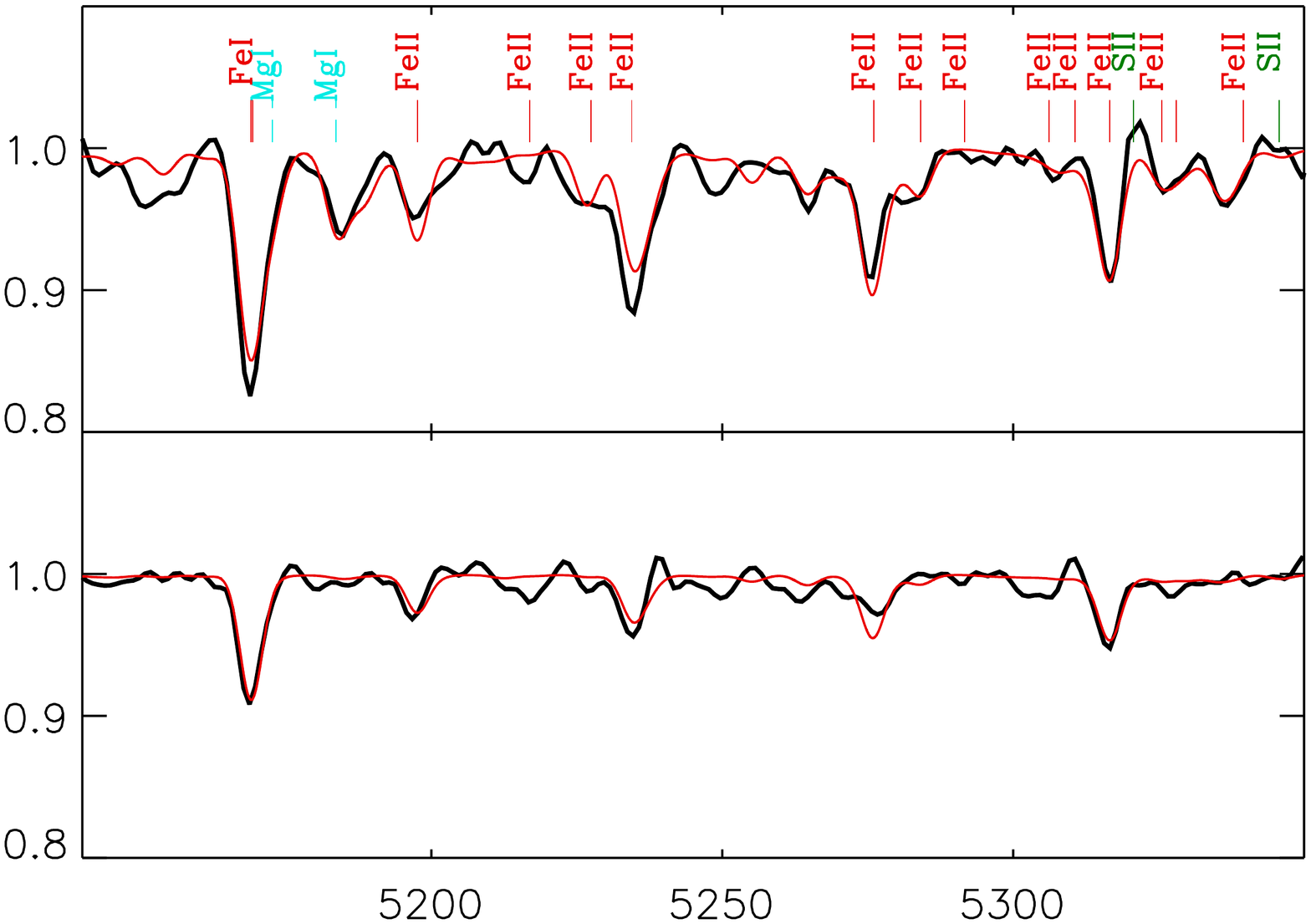}
\caption{\label{zfit1} Observed metal line spectra of stars C1 49 (top) and A 38 (bottom) in two spectral windows compared with model calculations (red) obtained for the final stellar 
parameters of Table~\ref{table_asg}.}
\end{figure*}

\begin{figure}
\begin{center}
\includegraphics[scale=0.50,angle=90]{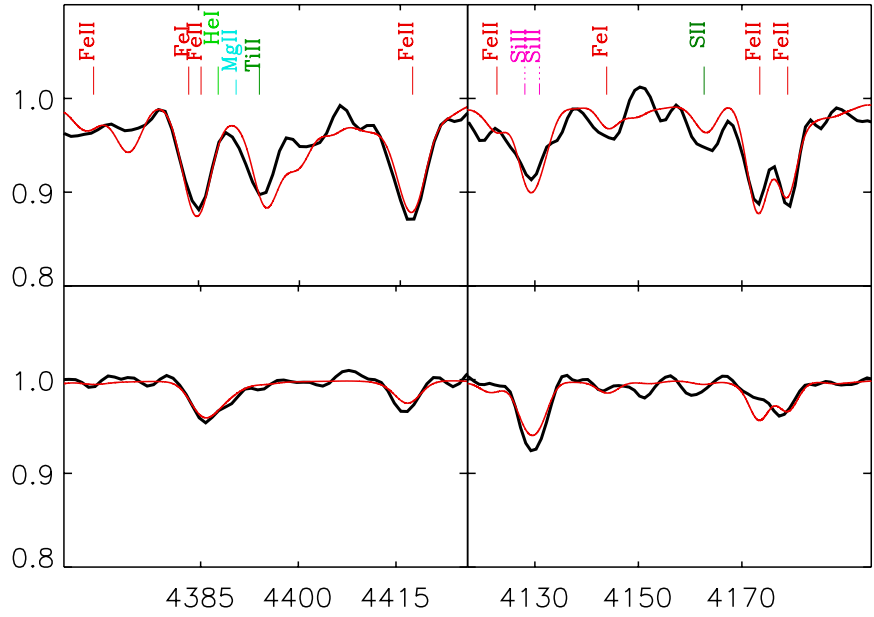}
\includegraphics[scale=0.50,angle=90]{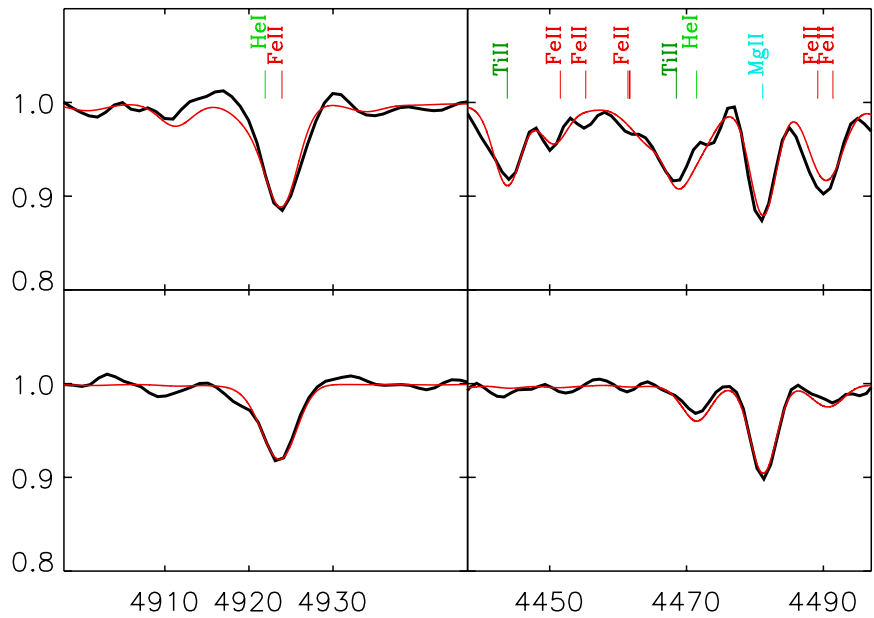}
\caption{\label{zfit2} Same as Fig.~\ref{zfit1} for four more spectral windows.}
 \end{center}
\end{figure}

Fig.~\ref{zfit1} and ~\ref{zfit2} give an impression of the metal line fits in different spectral windows. 
Generally, the fits of the spectra are very good and the stellar parameters are well constrained. Most importantly, the accuracy of the metallicities is about 0.1 dex. We note {\bf again} that the
metallicity [Z] derived here comes from a fit of metal lines over a wide range of elements, mostly iron, titanium, chromium, magnesium and silicon. In this sense, it represents an average over many
elements including the iron group. This is different from HII region metallicities where usually only oxygen is used as a proxy for metallicity. For the comparison of our stellar
with HII region metallicities we convert the latter to [Z] = 12 + log(O/H) $-$ 8.69, where 8.69 corresponds to the solar oxygen abundance \citep{asplund2009}. Such a comparison 
assumes that the ratios of $\alpha$-to-iron elements are solar.

\subsection{Evolutionary status}

\begin{figure*}
\includegraphics[scale=1.2]{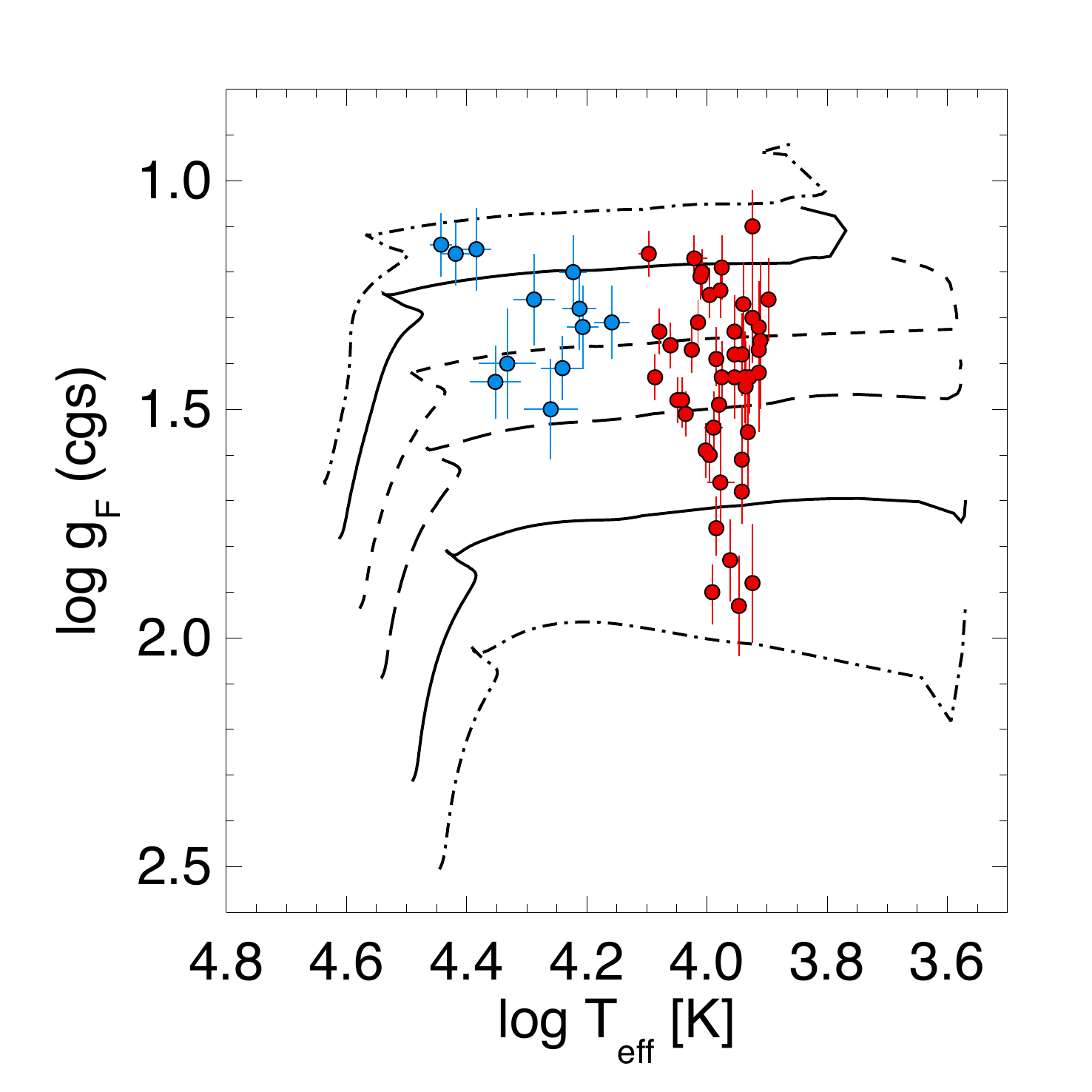}
\caption{\label{shrd} Spectroscopic Hertzsprung-Russell diagram of the NGC\,55 supergiant stars of this study compared with evolutionary tracks 
from \cite{eckstroem2012}. Blue: early B spectral type, red: late B and early A spectral type. The evolutionary tracks are calculated for initial main sequence masses of (from the bottom of the 
figure to top) 12, 15, 20, 25, 32 and 40 M$_{\odot}$, respectively, and include the effects of stellar rotation.}
\end{figure*}

The evolutionary status of the objects of our full sample can be assessed from Fig.~\ref{shrd}, which shows flux-weigthed gravities against effective temperatures compared with
stellar evolution calculations. As discussed in detail by \citet{langer2014}, this ``Spectroscopic Hertzsprung-Russell Diagram'' is a very useful tool to constrain  stellar properties 
in situations where either the distance is unknown or accurate photometry, which can be used to determine luminosity or bolometric magnitudes, is lacking.
From Fig.~\ref{shrd}, which compares with the evolutionary tracks by \citet{eckstroem2012}, it is apparent that the stars of our sample are in an advanced stage of stellar evolution. 
From their flux weighted gravities we conclude
that they represent a mass range from 15 to 40 M$_{\odot}$. Our sample of selected stars is, thus, comparable to those of our previous studies on other galaxies refererred to in the publications
already mentioned. The fact that no lower mass early B-type supergiants are present in our sample is a consequence of the lower $V$-magnitude limit in the selection of our targets for 
multi-object spectroscopy. Early B-type supergiants have a much higher bolometric correction than later spectral types. As a consequence, at the same level of $V$ magnitude these objects have
higher luminosities and, thus, lower flux-weighted gravities. 

\subsection{Metallicity and metallicity gradient}

\begin{figure}
\begin{center}
\includegraphics[scale=01.10]{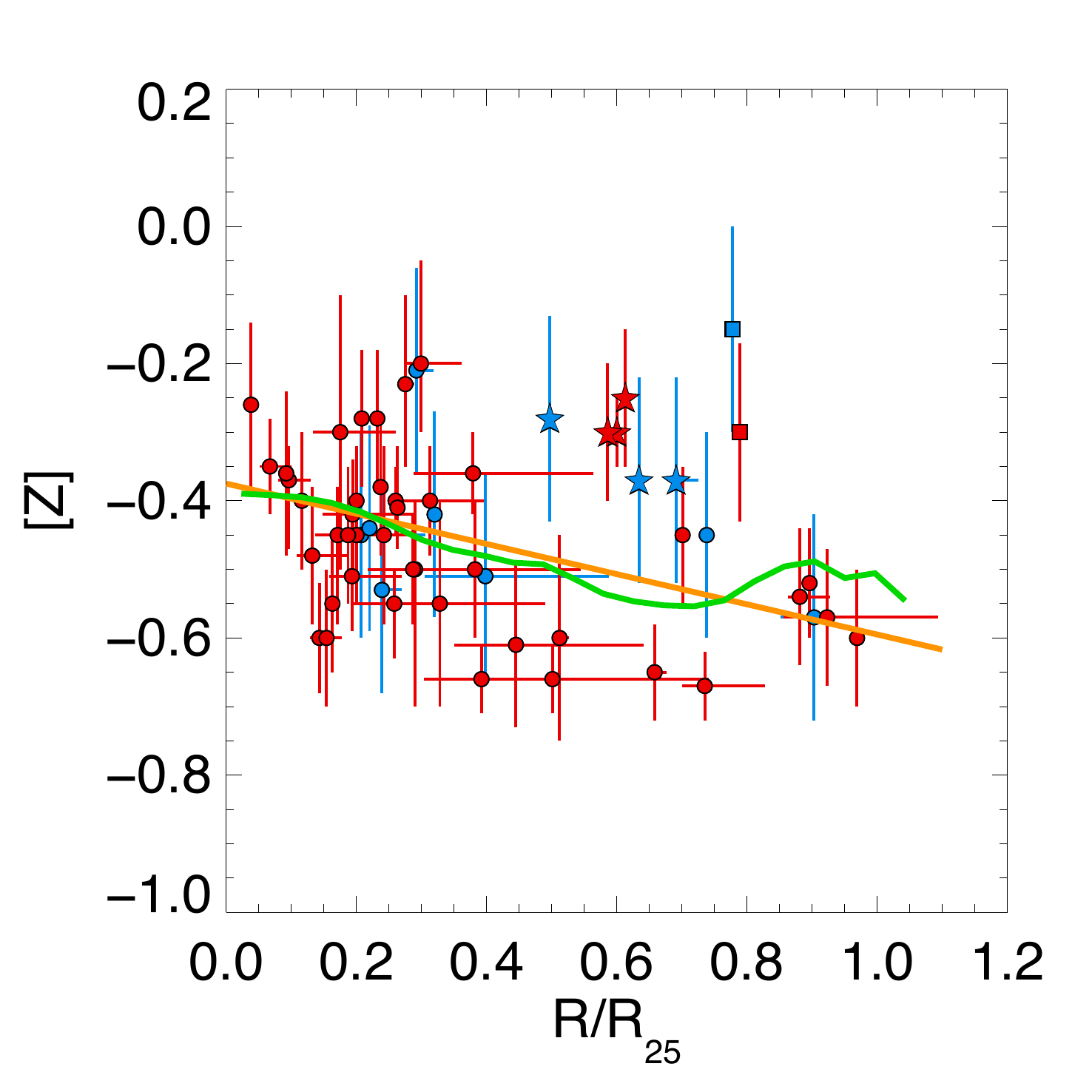}
\caption{\label{zgrad} Metallicity of the NGC\,55 blue supergiants as a function of galactocentric distance. Blue: early B spectral type, red: late B and early A spectral type. The linear regression
accounting for errors in metallicity and galactocentric distance is plotted in orange. Potential outliers are highligthed as squares and stars. The radial metallicity distribution calculated with
our chemical evolution model discussed in section 4 is shown in green.}
 \end{center}
\end{figure}

In Fig.~\ref{zgrad} we plot stellar metallicities as a function of galactocentric distance. We see that the two different groups of our sample, early B-type supergiants and late B-type/A-type
supergiants give consistent metallicities. This is reassuring, because different types of model atmospheres and atomic models were used for the non-LTE line formation calculations and different
methodologies were used for the analysis. We conclude that systematic effects caused by these differences are small. 

The average  metallicity is somewhat lower than in the LMC and between [Z] = $-0.4$ to $-0.5$. A linear regression of the form

\begin{equation}
 [Z]\,=\,[Z]_0\,+\,grad[Z]\,{R \over R_{25}}  
\end{equation}

accounting for errors in both the x-axis and the y-axis reveals a significant metallicity gradient. We obtain [Z]$_0$ = $-0.37 \pm 0.03$ dex and grad[Z] = $-0.22 \pm 0.06$ dex/R$_{25}$.
This is the first detection of a metallicity gradient in the disk of this galaxy.

\begin{figure}
\begin{center}
\includegraphics[scale=0.80]{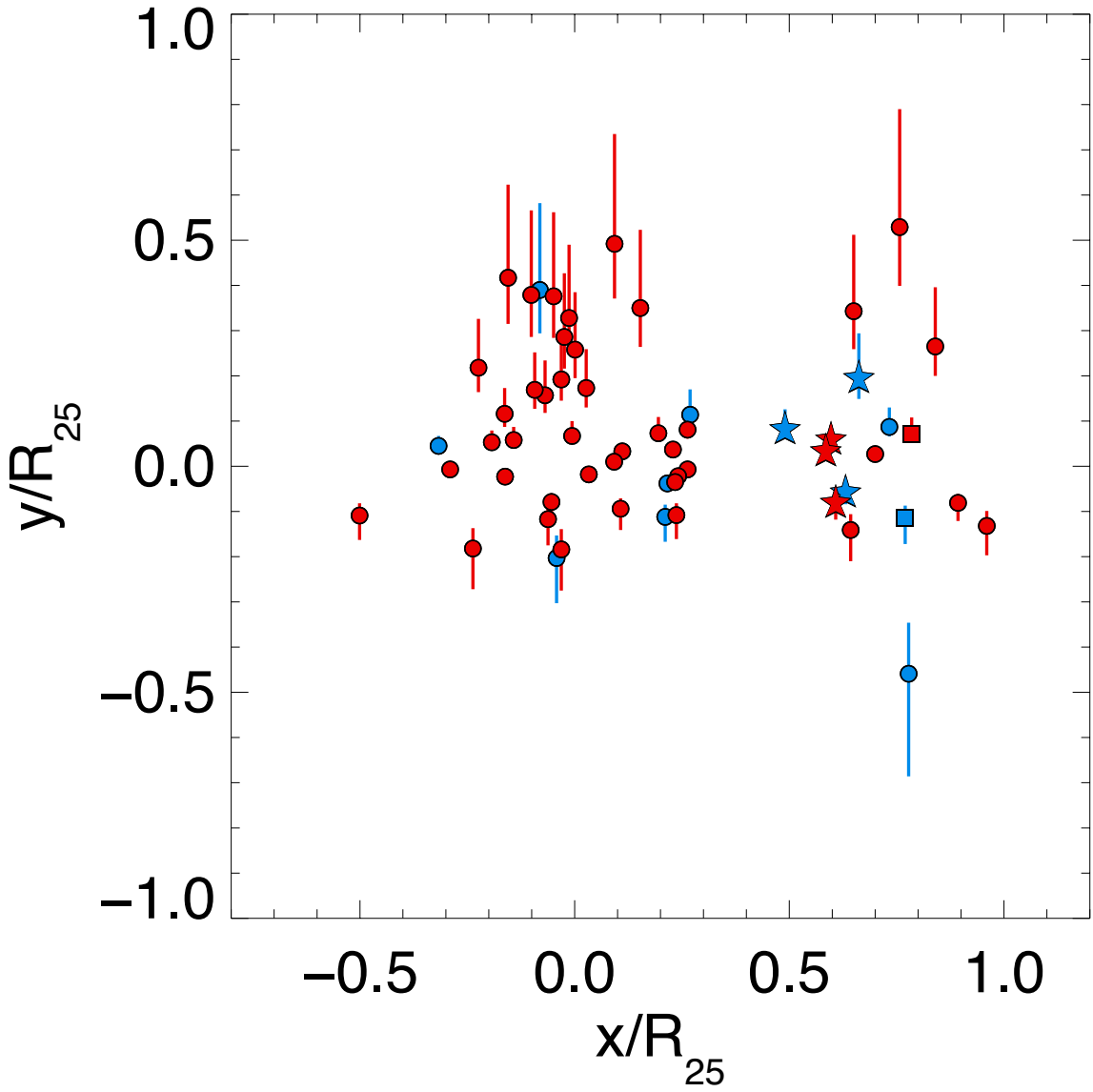}
\caption{\label{bsgloc} De-projected locations of the NGC\,55 blue supergiants in the galactic disk plane. Symbols are the same as in Fig.~\ref{zgrad}.}
 \end{center}
\end{figure}

The local scatter around this relation is about 0.15 dex. This is larger than most of the individual metallicity uncertainties and may be indicative of some chemical inhomogeneity in the
disk of this galaxy. A few objects above the regression between 0.5 and 0.8 R/R$_{25}$ seem to be outliers. We highlight them with different symbols (stars and squares) in Fig.~\ref{zgrad}. 
In order to test whether these outliers are spatially correlated we also plot the de-projected locations of our objects in Fig.~\ref{bsgloc}. We find indeed that these objects cluster 
on the major axis between 0.5 and 0.8 R/R$_{25}$. 

\begin{figure}
\begin{center}
\includegraphics[scale=0.80]{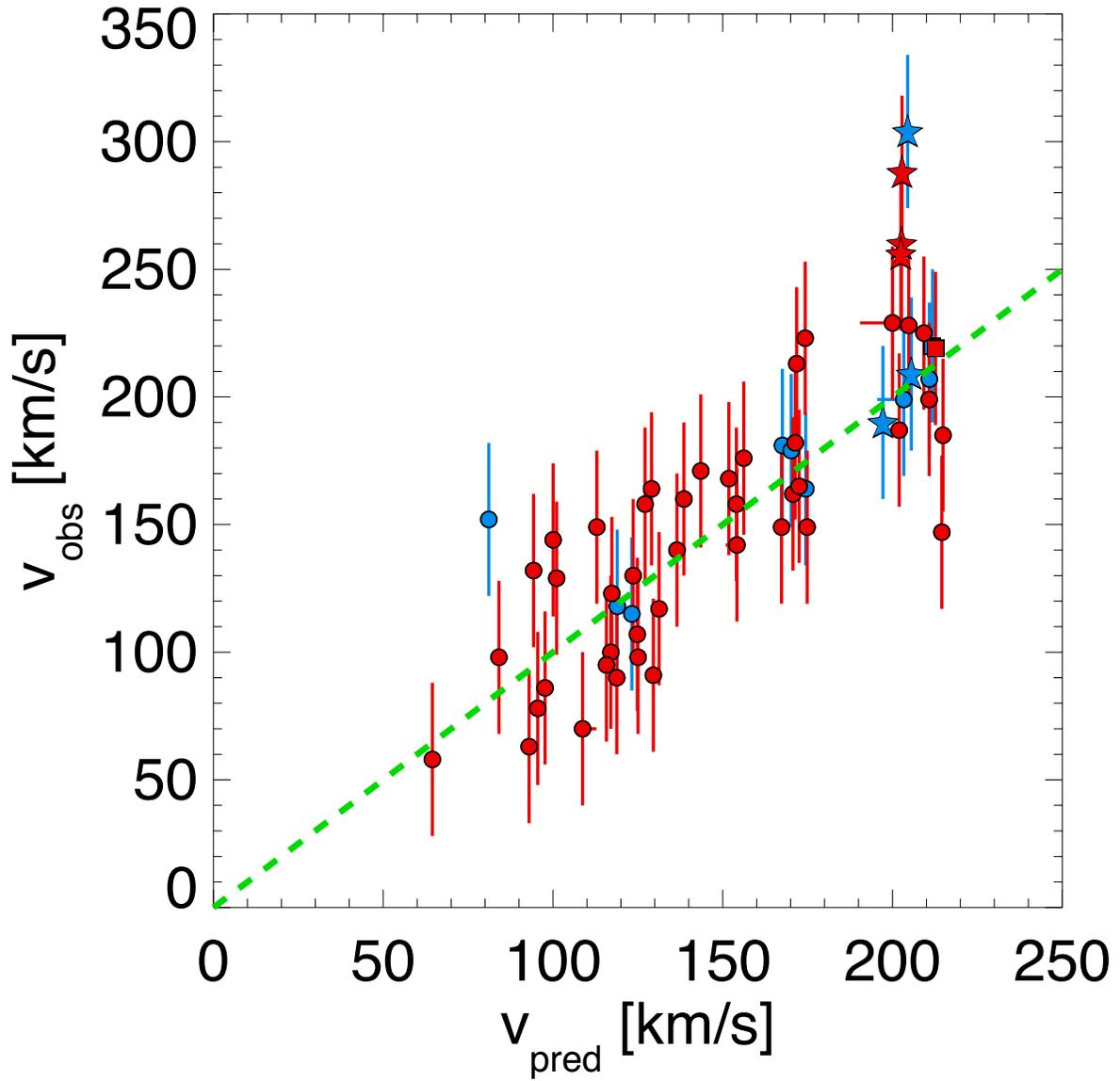}
\caption{\label{bsgvrad} Observed radial velocities of NGC\,55 blue supergiants compared with predicted velocities using the rotation curve determined by \citet{puche1991}. Symbols as in Fig.~\ref{zgrad}. The systemic velocity measured by \citet{westmeier2013} was adopted (see text).}
 \end{center}
\end{figure}

It is also interesting to check whether some of these chemical abundance outliers are dynamically distinct. For this purpose we use the rotation curve derived by \citet{puche1991} together with the 
new 
corrected systemic velocity of Table~\ref{table_asg} to calculate radial velocities of the objects of our sample predicted by the kinematic rotation model. The comparison with the observed radial
velocities measured by \citet{castro2008} is plotted in Fig.~\ref{bsgvrad}. As is evident from the figure, four of the eight suspected chemical outliers are definite kinematical outliers.

The level of chemical enrichment in the disk of NGC\,55 obtained from our blue supergiant analysis is generally in agreement with the oxygen abundances of the few HII regions studied. We have used the 
published line fluxes by \citet{webster1983} and applied the ``direct method'' using the auroral forbidden line [OIII]$\lambda$4363. We obtained [Z] = $-0.39$, $-0.34$, $-0.14$, $-0.36$ for their 
HII regions 1, 2, 3, 4, respectively. In the same way we find [Z] = $-0.52$, $-0.35$, $-0.11$ for the HII regions 3, 10, ``center'' 
investigated by \citet{stasinska1986}. For the central disk HII region in \citet{tuellmann2003} we obtain [Z] = $-0.52$.

We note that \citet{pilyugin2014}, applying their newly calibrated strong line method on eight HII regions in NGC\,55,
four of them in the center, obtain a value of $-0.65$ below solar and no significant abundance gradient. This is at odds with results obtained by \citet{webster1983} and \citet{castro2012}. It
is also not supported by our blue supergiant spectroscopy result. \citet{kudritzki2015} have already pointed to a systematic offset of 0.15 to 0.2 dex between the \citet{pilyugin2014} HII strong line calibration and their blue supergiant studies of a sample of galaxies. The difference encountered here goes into the same direction, but is slightly larger.  

\subsection{Mass-metallicity relationship}

\begin{figure}
\begin{center}
\includegraphics[scale=1.10]{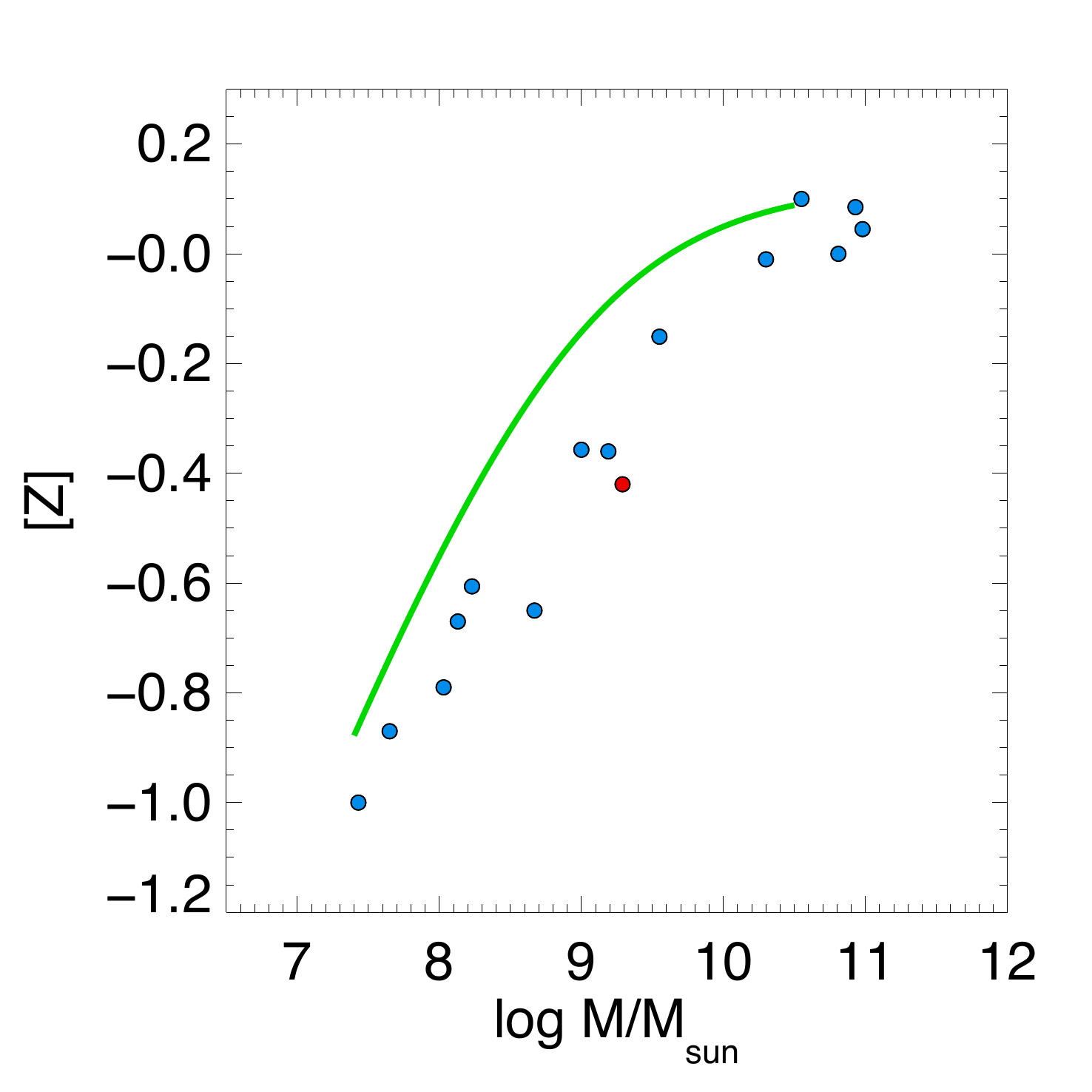}
\caption{\label{mzr} Relationship between total stellar mass and metallicity for a sample of 15 galaxies studied with supergiant spectroscopy (see text). NGC\,55 is indicated
in red. The green curve is the mass-metallicity relationship obtained by \citet{andrews2013} from HII region emission lines of stacked spectra of 50,000 SDSS galaxies using the direct method with 
auroral lines.}
 \end{center}
\end{figure}

The relationship between the total stellar mass of a galaxy and its average metallicity, the mass-metallicity relationship (``MZR''), is a Rosetta stone to understand the formation and evolution 
of galaxies. After the pioneering paper
by \citet{tremonti2004}, which studied 50,000 SDSS galaxies, revealed the existence of a tight MZR, a large effort has been made to theoretically interpret this relationship either by laborious
numerical simulations or by analytical models \citep[see, for instance,][]{zahid2014}. However, as shown by \cite{kewley2008} the original relationship based on the analysis of strong HII region 
emission lines is subject to large systematic errors, which are poorly understood. Therefore, \cite{kudritzki2012} have started to build up a MZR of galaxies in the local Universe, which is based 
solely on the results of stellar spectroscopy. Our analysis of the young stellar population in the disk of NGC\,55 now enables us to add an additional data point to this relationship. This is done
in Fig.~\ref{mzr}. Note that in order to be consistent with \citet{kudritzki2012} and \citet{kewley2008} we use the metallicity at two disk scale lengths (corresponding to 0.22 R/R$_{25}$ in 
the case of NGC\,55) for 
the plot. We have also added the new results by \citet{hosek2014} and \citet{kudritzki2014}. The stellar mass used for NGC\,55 is determined in the next section.

As is evident from Fig.~\ref{mzr} the stellar spectroscopy-based MZR does also form a tight relationship and NGC\,55 fits nicely into the middle of it. The SDSS MZR by \citet{andrews2013}, which
is based on stacked spectra of galaxies with similar mass so that the HII region auroral lines can be used for a more precise determination of oxygen abundances, shows a very similar shape.
However, there is small shift of 0.15 dex in metallicity or 0.3 dex in stellar mass. In view of the different techniques applied for the diagnostics of metallicities and stellar masses we find
the agreement compelling. It seems that many of the strong line calibrations discussed by \citet[see their Fig.~2]{kewley2008} can be ruled out. 

\subsection{The nature of the extra-planar HII regions in NGC\,55}

\citet{tuellmann2003} have detected two HII regions in the central region of NGC\,55 clearly above the plane of the galactic disk at heights of 0.8 and 1.5 kpc based on their adopted distance of 1.6 Mpc. With our new distance of 2.34 Mpc (see Section 5.) the heights above the disk plane increase to 1.13 and 2.22 kpc, respectively. They argue that 
hydrodynamical considerations rule out an ejection scenario of the ionizing O stars with the surrounding HII region gas from the central disk. As a consequence, the ionizing massive stars
must have been born in-situ over the disk. In order to support their argument they determine the metallicity of the two extra-planar HII regions and obtain [Z] = $-0.9 \pm 0.2$, a significantly
sub-solar value. They then study one central HII region in the disk and derive [Z] = $-0.64 \pm 0.1$. The abundances are derived using the auroral line [OIII]$\lambda$4363 (or an upper limit for the flux in
this line) and the R$_{23}$ method in the calibration by \citet{mcgaugh1991}. From the difference in metallicity they conclude that the extra-planar objects are
chemically distinct from the disk population, which supports the idea of in-situ formation over the disk. 

However, taking the uncertainties of the metallicties into account one could argue that at the margin of the errors they almost overlap. The chemical proof of in-situ formation all hinges on
the metallicity of the one central HII region studied. With our stellar study using a large sample of objects the central metallicity in the disk of NGC\,55 is accurately restricted to
[Z]$_0$ = $-0.37 \pm 0.03$ dex. This is a significant difference relative to the metallicity of the two extra-planar HII regions and provides very strong support to the scenario proposed by
\citet{tuellmann2003} of in-situ formation of massive stars above the disk, facilitated by the presence of large amounts of extraplanar gas due to pronounced mass outflow and inflow in this galaxy (see next Section).

\begin{deluxetable}{ c c c c c c }
\tablecaption{Spectral type - \teff\ relationships of blue supergiant stars\label{table_spteff}}
\tablehead{
\colhead{ }        & \colhead{this work} & \colhead{FP12}  & \colhead{K03}   & \colhead{EH03}  & \colhead{EH03}  \\
\colhead{[Z]}      & \colhead{NGC\,55}   & \colhead{MW}    & \colhead{MW}    & \colhead{MW}    & \colhead{SMC}   \\
\colhead{spectral} & \colhead{\teff}     & \colhead{\teff} & \colhead{\teff} & \colhead{\teff} & \colhead{\teff} \\
\colhead{type}     & \colhead{K}         & \colhead{K}     & \colhead{K}     & \colhead{K}     & \colhead{K}     \\[1mm]
\colhead{(1)}      & \colhead{(2)}	 & \colhead{(3)}   & \colhead{(4)}   & \colhead{(5)}   & \colhead{(6)}
}
\startdata
 B8    & 11800$\pm$490 & 12200$\pm$410 & 12000 & 13000 & 12000 \\
 B9    & 10600$\pm$330 & 10920$\pm$220 & 10500 & 10750 & 10500 \\
 A0    &  9800$\pm$340 &  9840$\pm$290 &  9500 &  9750 &  9500 \\
 A2    &  9000$\pm$330 &  8960$\pm$200 &  9000 &  9000 &  8500 \\
 A3    &  8600$\pm$150 &  8430$\pm$60  &  8500 &  8500 &  8000 \\
 A5    &  8130$\pm$130 &               &       &  8250 &  7750 \\
\enddata
\tablecomments{FP12: \citet{firnstein2012}; K03: \citet{kudritzki2003}; EH03: \citet{evans2003}.
}
\end{deluxetable}

\subsection{Relationship between spectral type and effective temperature}

The set of homogeneous spectra of 46 objects of spectral type B8 to A5 with well determined stellar parameters offers the opportunity to investigate the relationship between spectral type 
and effective temperature at a metallicity slightly lower than the LMC. For this purpose we use the spectral types assigned by \citet{castro2008} and the \teff\ values of our analysis provided in
in Table~\ref{table_asg}.  Table~\ref{table_spteff} gives the mean \teff\ and standard deviation for each subtype as determined by our spectral analysis. (At spectral type A2 two outliers, D 39 and 
C2 22, were encountered, which are not included in the calculation of the mean). The comparison with the relationships for the Milky Way obtained by \citet{kudritzki2003} and more recently by
\citet{firnstein2012} based on the detailed NLTE analysis of high resolution, high S/N spectra shows good agreement. The \citet{evans2003} relationship for the Milky Way is also in agreement, 
whereas their
relationship for the SMC is significantly cooler at later spectral types. We note that \citet{evans2003} have adopted a metallicity [Z] = -0.76 for the SMC, when they calibrated their relationship.
This metallicity is significantly lower than for most of our objects studied in NGC\,55. In addition, their calibration did not include NLTE effects for the calculation of the metal lines. LTE is
very likely sufficient for their SMC objects, because most of them are of luminosity class II. On the other hand, most of the NGC\,55 objects have higher luminosities corresponding to luminosity 
class I and require NLTE calculations to fit the spectrum. 

We conclude that at the metallicity level of NGC\,55 and for luminosity class I objects we do not see a systematic difference of the spectral type - relationship when compared with the Milky Way.

\section{A chemical evolution model for the disk of NGC\,55}

The stratification of metallicity of the young stellar population over the disk of a spiral galaxy compared with the radial profiles of ISM gas mass and stellar mass contains crucial information
about the chemical evolution history of the galactic disk. In the simple
closed box model, which neglects the effects of accretion inflow of material from the cosmic web and of outflow by galactic winds, metallicity at a certain galactocentric radius 
is uniquley determined by the ratio of the column densities of stellar mass and gas mass at this radius. Metallicity gradients are then a consequence of the variation of this ratio as function of 
of galactocentric radius. However, it is now well known that accretion and galactic winds play an important role in the chemical evolution of galaxies and, thus, the closed box model can only
provide qualitative insight into galaxy evolution. As a simple step further, \citet{kudritzki2015} have developed a chemical evolution model, in which accretion and outflow are taken
into account, but the assumption is made that the rates of mass-gain and mass-loss in units of the star formation rate have been constant. They provide qualitative arguments in their paper in support
of this simplifying assumption. Under this assumption, the equations of chemical evolution can be solved analytically and the observed radial metallicity and mass distributions can then be
used to constrain the rates of mass infall $\Lambda$ and outflow $\eta$ in units of the star formation rate $\psi$

 \begin{equation}\label{eta}
\eta \equiv {\dot{M}_{loss} \over \psi}
\end{equation}
and the mass accretion factor 
\begin{equation}\label{lam}
\Lambda \equiv {\dot{M}_{accr} \over \psi}.
\end{equation}

In this way, \citet{kudritzki2015} have determined mass-loss and accretion rates for 20 nearby disk galaxies. With a well constrained determination of central metallicity and
the metallity gradient it is an obvious step to apply the same analysis to NGC\,55. This is the first application of this approach on a galaxy with a well determined spatial 
distribution of metallicity of the young stellar population through spectroscopy. Similar to \citet{kudritzki2015} we obtain stellar mass column density profiles as a function of
galactocentric radius from de-projected mid-IR images at 3.4 $\mu$m obtained with the WISE (W1-band) and Spitzer (IRAC) space observatories (for details on the conversion from mid-IR
surface photometry to mass-column densities we refer to Kudritzki et al.). The ISM atomic gas mass column density distribution can be obtained from de-projected radio 21 cm observations, 
where the radio fluxes in the 21 cm line are converted into HI masses and then multiplied by a factor of 1.36 to include the mass of helium and metals. We have re-analyzed the original
VLA data cube obtained by \citet{puche1991} for that purpose. \citet{westmeier2013} find that the galaxies 21 cm flux is higher than the one obtained by \citet{puche1991}, which is very likely
caused by differences in the flux calibration or missing flux through poor coverage of the uv-plane. We apply a factor 1.3 to the Puche et al. data to correct for this effect.

In the inner disks of spiral galaxies the molecular gas can make an important contribution to the total gas mass. An estimate of the molecular gas mass is usually obtained from an observation
of CO which is then converted to molecular hydrogen mass, for which the factor 1.36 is then applied again to include helium and metals. Unfortunately, in the case of NGC\,55 no CO observations 
are available.
We, therefore, use an indirect way to estimate the column density distribution of the molecular mass. We start from the de-projected observation of the 21 cm continuum radio flux, which is then
converted into far-IR flux using the FIR-radio flux correlation and yielding the radial profile of star formation rate (see \citealt{ho2010}). The star formation rates are then transformed into 
molecular hydrogen column densities by application of the molecular Kennicutt-Schmidt law as parameterized by \citet{leroy2013}. 

The integrated total stellar mass of the disk of NGC\,55  (in units of the solar mass) is log M$_{star}$ = 9.29. We note that the mass depends on
the distance assumed. We have used a distance of 2.34~Mpc, which we derive in the following section.

\begin{figure}
\begin{center}
\includegraphics[scale=1.10]{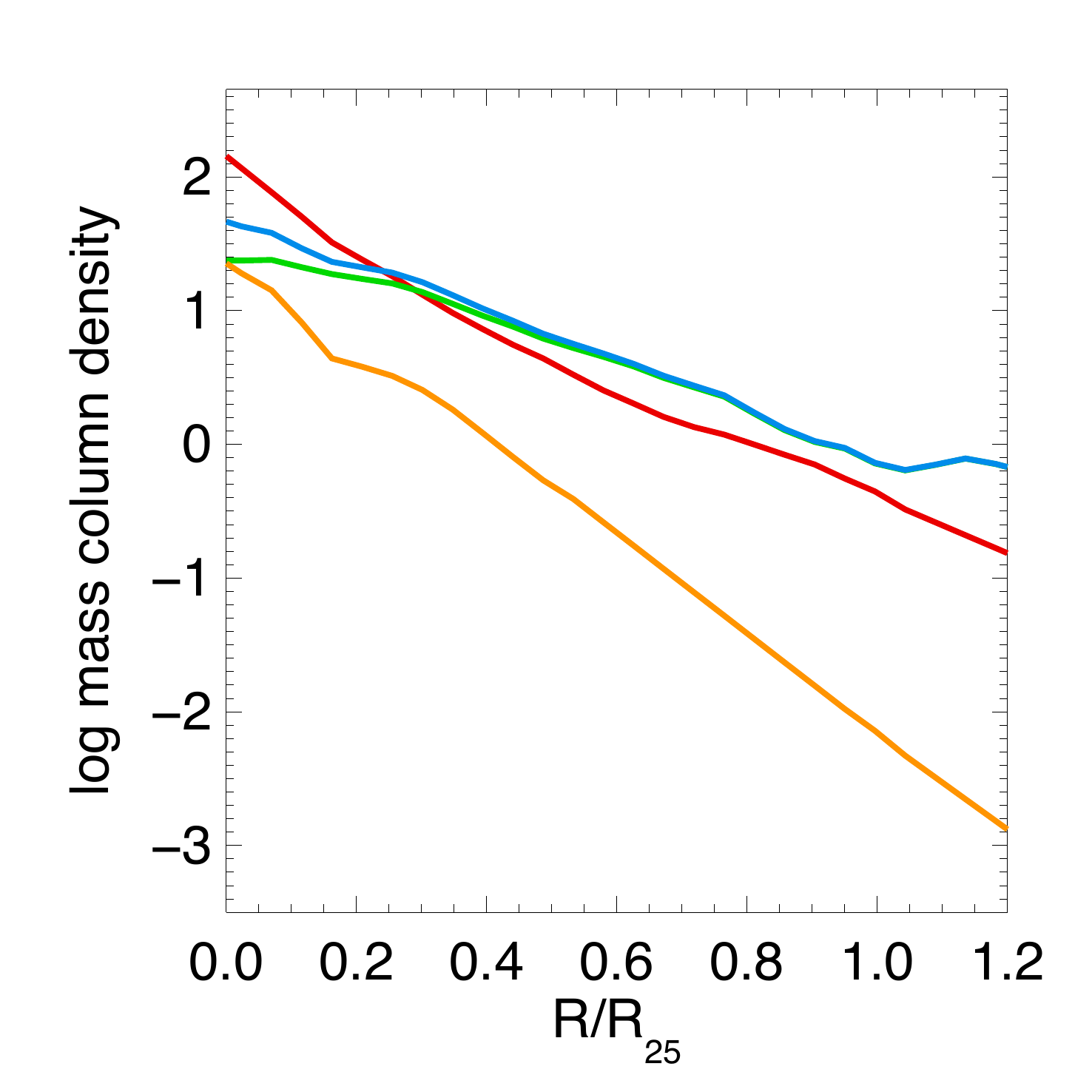}
\caption{\label{masses} Radial distribution of the logarithm of the de-projected NGC\,55 disk mass column densities (in M$_{\odot}/pc^2$). Red: stellar mass, blue: ISM gas mass, green: atomic gas mass
contribution to ISM, orange: molecular gas mass contribution to ISM.}
 \end{center}
\end{figure}

Fig.~\ref{masses} displays the column densities of stellar and ISM atomic and molecular masses obtained in this way as a function of galactocentric radius. Obviously, the molecular
gas contribution is only important in the central disk region and has only a weak effect on the constraints for outflow and accretion. The central stellar mass profile does not show a significant
indication of a bulge contribution. We, thus, refrain from an attempt of bulge de-composition as carried out for most of the galaxies in Kudritzki et al.

Applying the same fit algorithm of the observed metallicity distribution as Kudritzki et al. we can then constrain the rate of outflow and accretion. We obtain $\eta$ = 0.72 and $\Lambda$ = 1.38.
This means that the chemical evolution of NGC\,55 is characterized by large amounts of gas outflow and infall. Compared to all the other galaxies studied by Kudritzki et al. the infall rate of
NGC\,55 is by far the largest. Qualitatively, this agrees well with the detailed morphological, kinematic and dynamic study by \citet{westmeier2013} and their conclusion that ``internal and
external processes, such as satellite accretion or gas outflows, have stirred up the gas disc''.

\begin{figure}
\begin{center}
\includegraphics[scale=1.10]{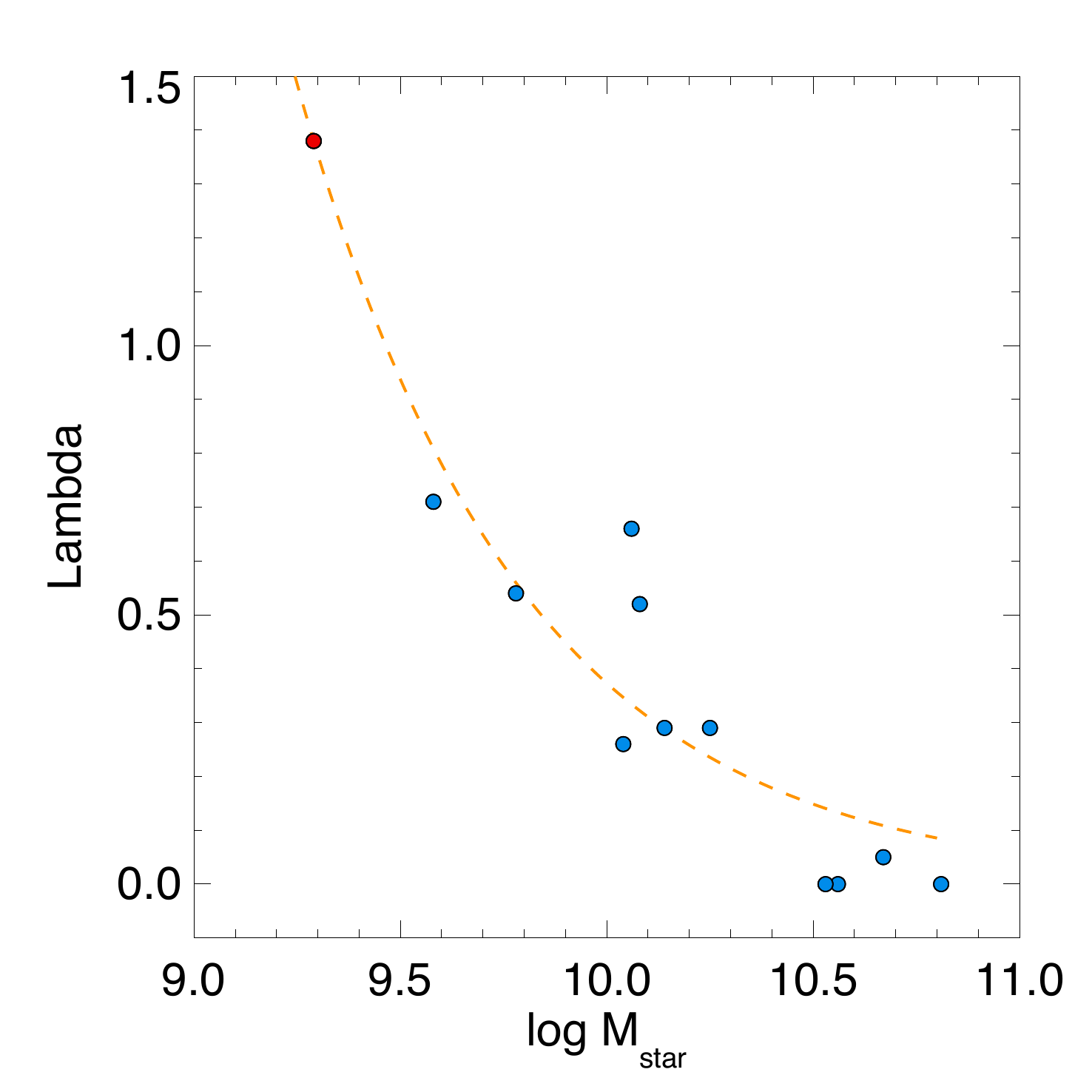}
\caption{\label{lambda} Galaxy mass accretetion rate $\Lambda$ measured in units of star formation rate as a function of the logarithm of integrated stellar mass (in solar units). Red: NGC\,55,
blue: galaxies studied by \citet{kudritzki2015} with well determined values of $\Lambda$ (see text). The orange dashed curve shows a power law fit $\Lambda \sim M_{star}^{-0.8}$.
}
 \end{center}
\end{figure}

With the large value of $\Lambda$ detected it is interesting to investigate in which way NGC\,55 is different from the other galaxies analyzed with the same chemical evolution approach.
We compare with the sample of galaxies with well determined $\eta$ and $\Lambda$ in Kudritzki et al. (see their Figure 11) and find the striking anti-correlation of $\Lambda$
with total galaxy stellar mass shown in Fig.~\ref{lambda}. From all the spiral galaxies studied so far with our method NGC\,55 has by far the lowest mass and it seems that $\Lambda$ increases 
significantly with decreasing mass. A power law $\Lambda \sim M_{star}^{-\alpha}$ with $\alpha=0.8$ provides a reasonable fit to the data, except at the high mass end, where very small values of
$\Lambda$ are observed. We note that on the ``main sequence of galaxies'' star formation increases with stellar mass following a power law $\psi \sim M_{star}^{\beta}$ with $\beta$
between 0.7 to 1.0 (see \citealt{lee2015} and references therein). Consequently, for the sample displayed in Fig.~\ref{lambda} the mass accretion rate $\dot{M}_{accr}$ can be expressed as
$\dot{M}_{accr} \sim M_{star}^{-\alpha+\beta}$. This means that the mass accretion rates depend only very weakly on galaxy mass, because $\alpha$ and $\beta$ almost cancel each other in the 
exponent. (At high galaxy masses accretion is probably reduced by shocks formed by the accretion process, which heat the infalling material and slow down accretion onto the disks).

To confirm this scenario is beyond the scope of this paper. We will need more studies in particular at the low mass end of the observed relation. 
This will require more detailed spatially resolved observations of the atomic and molecular gas in addition to multi-object stellar or HII-region spectroscopy and a careful investigations of 
star formation rates. The corresponding work is presently under way.

Fig.~\ref{zgrad} shows the metallicity distribution calculated with our best fit chemical evolution model compared to the observed metallicities. The model provides a good fit to the observed data.
We take this as a confirmation that the concept to retrieve important information on galaxy evolution from the spatial distribution of the metallicity of the young stellar population or the ISM
is a promising approach with large potential.
 
\clearpage
\begin{deluxetable}{ c c c c c c c c }
\tablecaption{NGC\,55 supergiants with HST photometry used for FGLR\label{table_asg_hst}}
\tablehead{
\colhead{Star} & \colhead{\logg$_F$} & \colhead{m$_{bol}$} & \colhead{V}    & \colhead{I}   & \colhead{E(B-V)} & \colhead{BC}  & (V-I)$_0$  \\
\colhead{ }    & \colhead{cgs}       & \colhead{mag}      & \colhead{mag}  & \colhead{mag} & \colhead{mag}    & \colhead{mag} & \colhead{mag} \\[1mm]
\colhead{(1)}  & \colhead{(2)}	     & \colhead{(3)}      & \colhead{(4)}  & \colhead{(5)} & \colhead{(6)}    & \colhead{(7)} & \colhead{(8)}
}
\startdata
 C1 17    & 1.43$\rm^{+0.08}_{-0.07}$ & 18.905$\pm$0.038 & 19.295$\pm$0.002 & 19.072$\pm$0.003 & 0.109$\pm$0.010 & -0.041$\pm$0.020 & +0.083$\pm0.012$ \\
 C2 15    & 1.59$\rm^{+0.06}_{-0.06}$ & 19.024$\pm$0.048 & 19.709$\pm$0.003 & 19.546$\pm$0.003 & 0.116$\pm$0.006 & -0.314$\pm$0.044 & +0.014$\pm0.006$ \\
 C2 13    & 1.43$\rm^{+0.09}_{-0.09}$ & 18.768$\pm$0.059 & 19.085$\pm$0.002 & 18.936$\pm$0.002 & 0.061$\pm$0.011 & -0.123$\pm$0.046 & +0.071$\pm0.014$ \\
 C1 14    & 1.32$\rm^{+0.10}_{-0.10}$ & 18.551$\pm$0.068 & 18.790$\pm$0.002 & 18.584$\pm$0.002 & 0.077$\pm$0.018 & +0.008$\pm$0.037 & +0.107$\pm0.008$ \\
 C1 34    & 1.36$\rm^{+0.05}_{-0.05}$ & 18.155$\pm$0.065 & 19.484$\pm$0.003 & 19.204$\pm$0.003 & 0.229$\pm$0.007 & -0.598$\pm$0.062 & -0.013$\pm0.008$ \\
 C2 37    & 1.54$\rm^{+0.07}_{-0.08}$ & 18.864$\pm$0.075 & 19.534$\pm$0.003 & 19.338$\pm$0.003 & 0.129$\pm$0.009 & -0.256$\pm$0.069 & +0.030$\pm0.010$ \\
 C2 29    & 1.16$\rm^{+0.05}_{-0.05}$ & 17.417$\pm$0.100 & 19.239$\pm$0.002 & 18.863$\pm$0.003 & 0.314$\pm$0.012 & -0.819$\pm$0.092 & -0.026$\pm0.015$ \\
 C2 22$^{\tablenotemark{a}}$    & 1.17$\rm^{+0.05}_{-0.05}$ & 17.394$\pm$0.107 & 18.737$\pm$0.002 & 18.336$\pm$0.002 & 0.277$\pm$0.015 & -0.456$\pm$0.096 & +0.046$\pm0.019$ \\
 C2 19    & 1.61$\rm^{+0.12}_{-0.12}$ & 19.532$\pm$0.061 & 19.532$\pm$0.003 & 19.729$\pm$0.003 & 0.099$\pm$0.014 & -0.064$\pm$0.042 & +0.056$\pm0.017$ \\
 C1 16    & 1.37$\rm^{+0.11}_{-0.11}$ & 18.872$\pm$0.051 & 19.281$\pm$0.002 & 19.025$\pm$0.002 & 0.131$\pm$0.011 & +0.010$\pm$0.037 & +0.088$\pm0.014$ \\
 C1 11$^{\tablenotemark{a}}$   & 1.45$\rm^{+0.11}_{-0.10}$ & 19.418$\pm$0.058 & 19.848$\pm$0.003 & 19.621$\pm$0.003 & 0.116$\pm$0.012 & -0.058$\pm$0.042 & +0.078$\pm0.015$ \\
 C2 14    & 1.33$\rm^{+0.05}_{-0.05}$ & 18.554$\pm$0.047 & 19.328$\pm$0.002 & 19.321$\pm$0.003 & 0.027$\pm$0.005 & -0.689$\pm$0.044 & -0.027$\pm0.005$ \\
 C1 13    & 1.15$\rm^{+0.09}_{-0.09}$ & 16.328$\pm$0.133 & 19.378$\pm$0.003 & 19.301$\pm$0.003 & 0.211$\pm$0.008 & -2.375$\pm$0.130 & -0.303$\pm0.014$ \\
\enddata
\tablecomments{Bolometric correction BC and intrinsic color (V-I)$_0$ calculated from model atmospheres (see \citealt{kudritzki2008} for details).
}
\tablenotetext{a}{Not included in distance determination because of variability (see text).}
\end{deluxetable}

\section{Distance}

The sub-sample of objects in Table~\ref{table_asg_hst} with HST/ACS photometry can be used to determine a distance using the FGLR technique briefly described in section 1. With
temperature, gravity and metallicity determined by spectroscopy we can calculate the intrinsic spectral energy distribution of our targets and their $V-I$ colors. Comparison with the observed colors then 
yields the reddening $E(B-V)$ [we apply the relation $E(B-V) = 0.78\,E(V-I)$] and, by assuming $R_V = A_V/E(B-V) = 3.2$, the value of extinction $A_V$. Adding the bolometric correction calculated
from the intrinsic energy distribution to the de-reddened $V$-magnitude we then obtain de-reddened apparent bolometric magnitudes for each object. The values obtained for the reddening, the bolometric
corrections and the bolometric magnitudes are also given in Table~\ref{table_asg_hst}, together with the logarithm of the flux-weighted gravities.

The NGC\,55 Cepheid search carried out by \citet{pietrzynski2006} in $V$ and $I$ provided us with repeated photometry of the targets in Table~\ref{table_asg_hst} over a period of 900 days, allowing us to
assess their photometric variability. According to \citet{bresolin2004} blue supergiants usually show only a small amount of photometric variability, which does not significantly affect
distances determinations through the FGLR method. But in a few cases larger photometric variations are encountered. Of the objects in Table~\ref{table_asg_hst}, star C2 22 showed clear signs of
variability, increasing in brightness in both $V$ and $I$  by 0.2 mag over the observed period. It was, therefore, not included in the final sample used for the distance determination. 
In addition, C1 11 showed brightness variations up to 0.1 mag and was also dropped from the sample. For all other objects the level of variability inferred from correlated changes in $V$ and $I$
is smaller than or equal to 0.05 mag. For those objects we add an additional uncertainty of 0.05 mag to the bolometric magnitude uncertainties in Table~\ref{table_asg_hst} to account for 
blue supergiant variability in the distance determination described below.

\begin{figure}
\begin{center}
\includegraphics[scale=1.10]{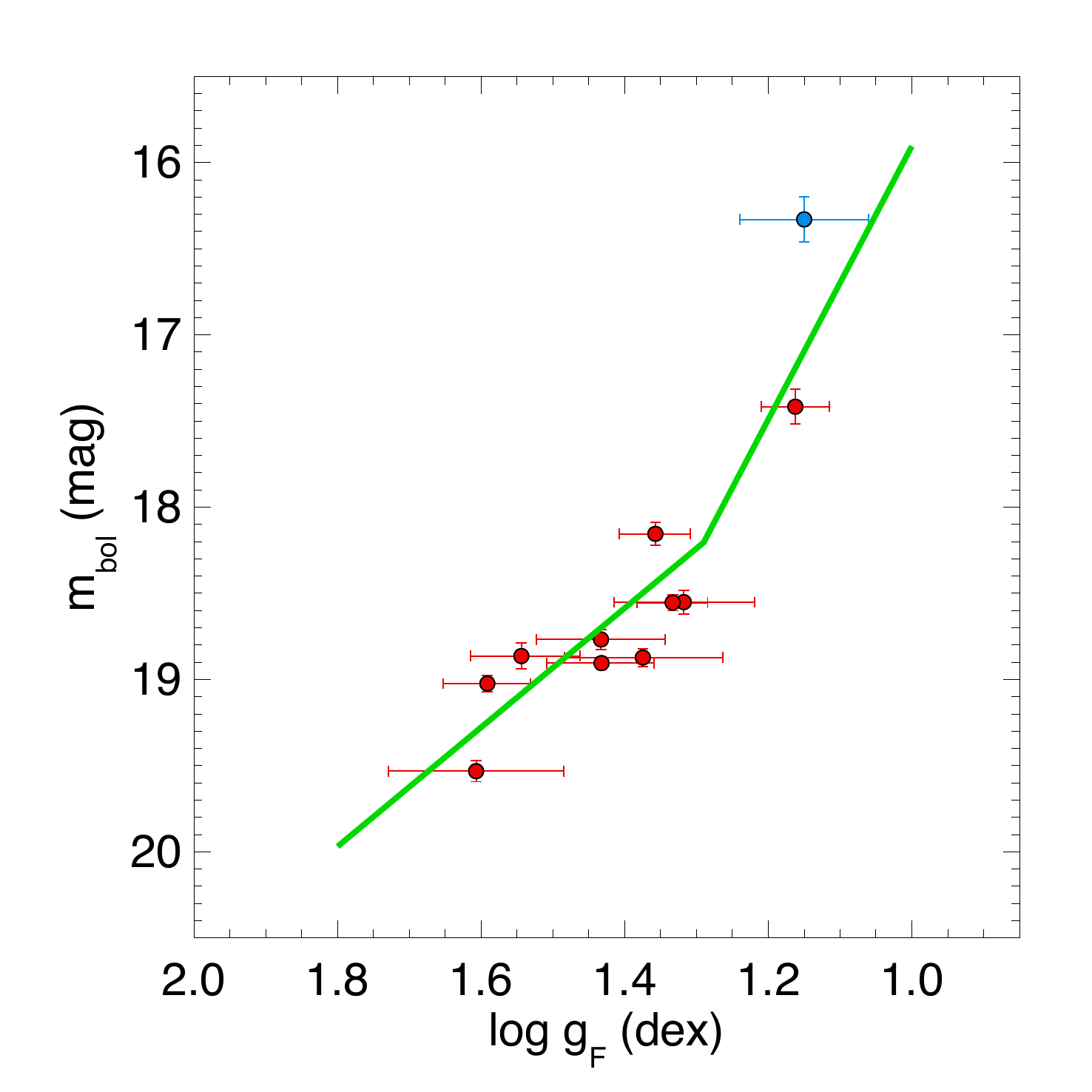}
\caption{\label{fglr} Observed FGLR for the sub-sample of NGC\,55 supergiants with HST photometry. The green relation is the LMC FGLR calibration \citep{urbaneja2016} shifted to the distance
modulus of $m-M = 26.80$ mag. Two targets of Table~\ref{table_asg_hst} are not included in the plot because of photometric variability (see text).}
 \end{center}
\end{figure}

Fig.~\ref{fglr} demonstrates that the bolometric magnitudes and flux-weighted gravities of our objects are tightly correlated. This is the ``Flux-weighted Gravity -- Luminosity Relationship (FGLR)''
introduced in section 1. As in previous work \citep{kudritzki2012, kudritzki2014, hosek2014, u2009, urbaneja2008} it can be used to determine extragalactic distances. Most recently, \citet{urbaneja2016}
have introduced a new calibration of the FGLR through a spectroscopic study of 90 blue supergiants in the LMC. Based on the analysis of the (M$_{bol}$, log g$_F$) diagram of their 90 LMC
supergiants they find that the FGLR changes its slope at log g$_F$ = 1.29. They fit a 2-component FGLR with two slopes to their data and determine a new zero point. The new FGLR in absolute
bolometric magnitude M$_{bol}$ is given by

\begin{equation}
 \log\,g_F\, \geq \log\,g_F^{\rm break}\,:\,   M_{\rm bol}\,=\,a (\log\,g_F\,-\,1.5)\,+\,b
\end{equation}
and 
\begin{equation}
\log\,g_F\, \leq \log\,g_F^{\rm break}\,:\,     M_{\rm bol}\,=\,a_{\rm low} (\log\,g_F\,-\,\log\,g_F^{\rm break}\,)\,+\,b_{\rm break}  
\end{equation}
with
\begin{equation}
 b_{\rm break}\,=\,a (\log\,g_F^{\rm break}\,-\,1.5)\,+\,b.
\end{equation}

with log g$_F^{\rm break} $ = 1.29, $a = 3.46 \pm 0.12$, $b = -7.92 \pm 0.03$ mag and $a_{\rm low} = 7.93 \pm 0.25$.

To determine the distance to NGC\,55 we use the observed flux-weighted gravities to calculate absolute magnitudes by means of the calibration FGLR. We than calculate individual distance
moduli for each object and obtain the final distance from a weighted mean which accounts for the observational errors in bolometric magnitude and the logarithm of flux-weighted gravity.
We obtain a distance modulus of $m-M = 26.85 \pm 0.08(r) \pm 0.05(s)$. The systematic error comes from uncertainties of the FGLR parameters. Fig.~\ref{fglr} also shows the calibration FGLR
shifted to the distance determined.

\begin{figure}
\begin{center}
\includegraphics[scale=1.10]{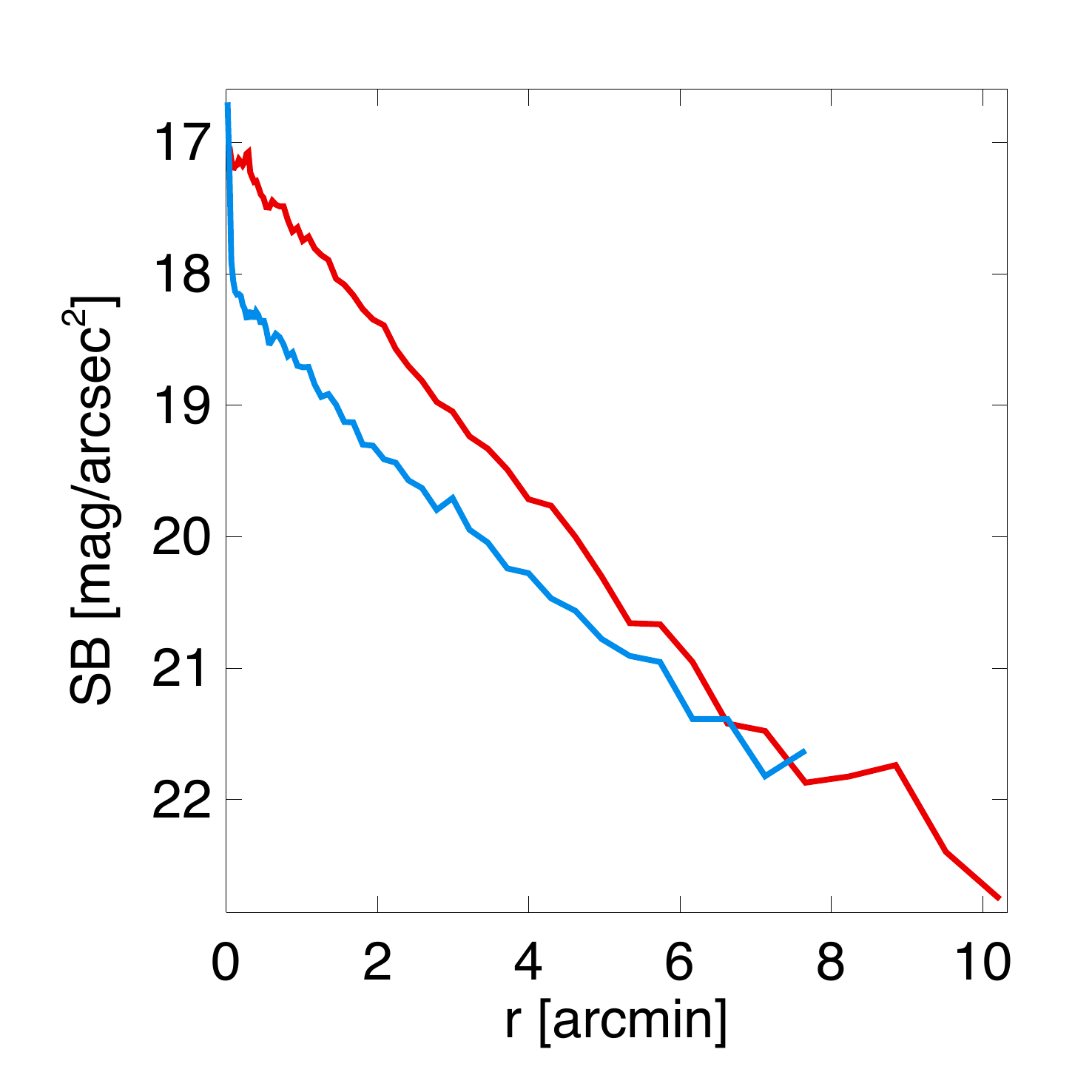}
\caption{\label{sfb} $K$-band radial surface brightness distribution of NGC\,55 (red) and and NGC\,300 (blue). Data from the 2MASS website (irsa.ipac.caltech.edu/applications/2MASS/LGA/).}
 \end{center}
\end{figure}

We realize that our FGLR-based distance modulus is 0.42 mag larger than the $ m-M =  26.43 \pm 0.09$ obtained by \citet{gieren2008} in our Araucaria collaboration. 
Gieren et al. use ground based $J$- and $K$-band photometry
of Cepheids in conjunction with the $V$ and $I$ photometry by \citet{pietrzynski2006}. This is a very reliable technique, in particular, because of the use of $K$-band, which minimizes the effects
of reddening for the Cepheids (we note that the average reddening of the targets in Table~\ref{table_asg_hst} is $E(B-V) \approx 0.15$ mag, very similar to the reddening found in the Cepheid study).
Usually the agreement between Cepheid and FGLR distances is satisfactory. For instance, in the case of the Sculptor group neighbor galaxy NGC\,300 our new FGLR calibration yields a distance
modulus of $m-M = {\bf 26.33}$, while \citet{gieren2005b} obtain 26.37 again from ground-based photometry combining the $V$, $I$, $J$ and $K$ bands. 
As mentioned in the Section 1., the physical parameters of the two galaxies are very similar. However, there is one fundamental difference between NGC\,300 and NGC\,55,
namely the fact that the latter is almost an edge on galaxy. As the result, the surface brightness of NGC\,55, for instance in the $K$-band, is significantly brighter than for NGC\,300 (see Fig.~\ref{sfb}).
This increases the potential of blending with nearby fainter objects, which would increase the measured brightness of the Cepheids and lead to a shorter distance. We recall that 
\citet{bresolin2005} have tested the effects of crowding for the case of NGC\,300 by comparing $V$- and $I$-band HST photometry with the ground-based observations and found only a small systematic
effect of 0.04 mag. However, in the case of NGC\,55 with a surface brightness a magnitude higher the systematic effect might be significantly larger. On the other hand, blue supergiants are several
magnitudes brighter than Cepheids and, thus, much less affected by blending. However, even for the supergiants we detect large differences between HST and ground based photometry in the case
of NGC\,55. We note that the TRGB distance to NGC\,55
is $26.62 \pm 0.1$ mag \citep[EDD database, http://edd.ifa.hawaii.edu, see][]{tully2009}. The planetary nebulae luminosity function (PLNF) distance modulus by \citet{vansteene2006} is 26.8 mag, albeit
with a relatively large error of $\pm 0.3$ mag. Should our FGLR distances be correct, then NGC\,55 would be at a slightly larger distance than NGC\,300, 2.34 Mpc versus 1.85 Mpc. 
With an angular separation of eight degrees \citep{westmeier2013} the physical separation would increase from 270 kpc (with the two galaxies at the same distance) to 570 kpc. 

At this point, the reasons for the discrepancy between the FGLR and Cepheid distances remain unresolved. A larger sample
of blue supergiants with HST photometry and a measurement of HST magnitudes for the Cepheids will be crucial for investigating this issue further.

\section{Conclusions}

The comprehensive spectroscopic study of blue supergiant stars distributed over the disk of NGC\,55 has led to the first detection of a metallicity gradient in this almost edge on late-type
spiral galaxy. The application of a chemical evolution model indicates the effects of intensive infall and outflow, in agreement with recent radio observations of the galaxy, which conclude that
the disk is very likely stirred up and disturbed by infalling and outflowing gas. We also find indications of chemical inhomogeneities, which support this picture. The significant difference
between the central metallicity of the disk and the metallicity of two extra-planar HII regions above the central disk provides strong evidence for in-situ star formation outside the galactic plane.
A distance determination using the FGLR method leads to a distance which is larger than the distance to the Sculptor group neighbor galaxy NGC\,300. Further HST photometry will be needed to
settle the issue of distance determination to this galaxy.

\acknowledgments

The authors wish to express their thanks to the referee whose critical and constructive remarks helped to improve the paper.
RPK and FB acknowledge support by the National Science Foundation under grants AST-1008789 and AST-1108906. RPK, MAU, WG
were supported by the Munich Institute for Astro- and Particle Physics (MIAPP) of the DFG 
cluster of excellence \textquotedblleft Origin and Structure of the Universe\textquotedblright. 
WG and GP gratefully acknowledge financial support for this work received from the BASAL Centro de Astrof\'{\i}sica y 
Tecnolog\'{\i}as Afines (CATA), PFB-06/2007. WG also acknowledges support from the Millenium Institute of Astrophysics (MAS)
of the Iniciativa Cient\'{\i}fica Milenio del Ministerio de Econom\'{\i}a, Fomento y Turismo de Chile, grant IC120009. In addition,
support from the Ideas Plus grant of the Polish Ministry of Science and Higher Education and TEAM subsidies of the Foundation for
Polish Science (FNP) is acknowledged by GP.
\vspace{5mm}
\facilities{VLT:FORS2, HST:ACS}


\clearpage

\clearpage

\clearpage

\clearpage

\clearpage

\clearpage

\clearpage

\clearpage

\clearpage

\clearpage

\clearpage

\clearpage

\clearpage


\clearpage


\clearpage 
\begin{thebibliography}{}

\bibitem[Andrews \& Martini(2013)]{andrews2013} Andrews, B.~H., Martini, P.\ 2013, \apj, 765, 140


\bibitem[Asplund et al.(2009)]{asplund2009} Asplund, M., Grevesse, N., Sauval, A.~J., \& Scott, P.\ 2009, \araa, 47, 481 

\bibitem[Bresolin et al.(2004)]{bresolin2004} Bresolin, F., Pietrzy\'nski, G., Gieren, W. et al.\ 2004, \apj, 600, 182

\bibitem[Bresolin et al.(2005)]{bresolin2005} Bresolin, F., Pietrzy\'nski, G., Gieren, W. et al.\ 2005, \apj, 634, 1020

\bibitem[Bresolin et al.(2009)]{bresolin2009} Bresolin, F., Gieren, W., Kudritzki, R.~P. et al.\ 2009, \apj, 700, 309





\bibitem[Castro et al.(2008)]{castro2008} Castro, N., Herrero, A., Garcia, M. et al.\ 2008, \aap, 485, 41

\bibitem[Castro et al.(2012)]{castro2012} Castro, N., Urbaneja, M.~A., Herrero, A.,net al.\ 2012, \aap, 542, A79

\bibitem[Cutri et al.(2003)]{cutri2003} Cutri, R.~M., Skrutskie, 
M.~F., van Dyk, S., et al.\ 2003, VizieR Online Data Catalog, 2246, 0 

\bibitem[Dalcanton et al.(2009)]{dalcanton2009} Dalcanton, J.~J., Williams, B.~F., Seth, A.~C. det al.\ 2009, \apjs, 186, 67

\bibitem[Dale et al.(2009)]{dale2009} Dale, D.~A., Cohen, S.~A., Johnson., L.~C. et al.\ 2009, \apj, 705, 514

\bibitem[de Vaucouleurs (1961)]{devaucouleurs1961} de Vaucouleurs, G.\ 1961, \apj, 133, 405

\bibitem[de Vaucouleurs et al.(1991)]{devaucouleurs1991} de Vaucouleurs, G., et al.\ 1991, Third Reference Catalogue of Bright galaxies, Volume I to III 


\bibitem[Eckstr{\"o}m et al.(2012)]{eckstroem2012} Eckstr{\"om}, S., Georgy, C., Eggenberger, P., et al.\ 2012, \aap, 537, A146

\bibitem[Evans \& Howarth(2003)]{evans2003} Evans, C.~J. \& Howarth, I.~D.\ 2003, \mnras, 345, 1223
 



\bibitem[Firnstein \& Przybilla(2012)]{firnstein2012} Firnstein, M., \& Przybilla, N.\ 2012, \aap, 543, AA80 




\bibitem[Gazak et al.(2015)]{gazak2015} Gazak, J.~Z., Kudritzki, R.~P., Davies, B. et al.\ 2015, \apj, 805, 182

\bibitem[Gieren et al.(2005a)]{gieren2005a} Gieren, W., et al.\ 2005a, ESO Messenger, 121, 23

\bibitem[Gieren et al.(2005b)]{gieren2005b} Gieren, W., Pietrzy\'nski, G., Soscynski, L. et al.\ 2005b, \apj, 628, 695

\bibitem[Gieren et al.(2008)]{gieren2008} Gieren, W., Pietrzy\'nski, G., Soscynski, L. et al.\ 2008, \apj, 672, 266



\bibitem[Ho et al.(2010)]{ho2010} Ho, I.-T., Wang, W.-H., Morrison, G.~E., Miller, N.~A.\ 2010, \apj, 722, 1051

\bibitem[Hosek et al.(2014)]{hosek2014} Hosek, M.~W., Jr., 
Kudritzki, R.-P., Bresolin, F., et al.\ 2014, \apj, 785, 151 


\bibitem[Hummel et al.(1986)]{hummel1986} Hummel, E., Dettmar, R.~J., \& Wielebinski, R.\ 1986, \aap, 166, 97

\bibitem[Humphreys (1988)]{humphreys1988} Humphreys, R.~M.,\ 1988, in The Extragalactic Distance Scale, ed. S. van den Bergh \& C.~J. Pritchet (Provo: Brigham Young Univ. Press), 103  

\bibitem[Jarrett et al.(2003)]{jarrett2003} Jarrett, D.~H., Chester, T., Cutri, R.\ 2003, \aj, 125, 525 

\bibitem[Jerjen et al.(1998)]{jerjen1998} Jerjen, H., Freeman, K.~C., \& Binggeli, B.\ 1998, \aj, 116, 2873 

\bibitem[Kang et al.(2016)]{kang2016} Kang, X., Zhang, F., Chang, R., et al.\ 2016, \aap, 585, 20

\bibitem[Karachentsev et al. (2003)]{karachentsev2003} Karachentsev, I.~D., Grebel, E.~K, Sharina, M.~E. et al.\ 2003, \aap, 404, 93

\bibitem[Kewley \& Ellison(2008)]{kewley2008} Kewley, L.~J., \& Ellison, S.~L.\ 2008, \apj, 681, 1183

\bibitem[Kudritzki \& Puls(2000)]{kudritzki2000} Kudritzki, R.~P., \& Puls, J.\ 2000, \araa, 38, 613 

\bibitem[Kudritzki et al.(2003)]{kudritzki2003} Kudritzki, R.~P., 
Bresolin, F., \& Przybilla, N.\ 2003, \apjl, 582, L83 

\bibitem[Kudritzki et al.(2008)]{kudritzki2008} Kudritzki, R.~P., 
Urbaneja, M.~A., Bresolin, F., et al.\ 2008, \apj, 681, 269 

\bibitem[Kudritzki et al.(2012)]{kudritzki2012} Kudritzki, R.~P., Urbaneja, M.~A., Gazak, Z., et al.\ 2012, \apj, 747, 15  

\bibitem[Kudritzki et al.(2013)]{kudritzki2013} Kudritzki, R.~P., Urbaneja, M.~A., Gazak, Z., et al.\ 2013, \apj, 779, L20  

\bibitem[Kudritzki et al.(2014)]{kudritzki2014} Kudritzki, R.~P., 
Urbaneja, M.~A., Bresolin, F., Hosek, M.~W., Jr., 
\& Przybilla, N.\ 2014, \apj, 788, 56 

\bibitem[Kudritzki et al.(2015)]{kudritzki2015} Kudritzki, R.~P., Ho, I.-T., Schruba, A., et al.\ 2015, \mnras, 450, 342


\bibitem[Langer \& Kudritzki(2014)]{langer2014} Langer, N., Kudritzki, R.~P.\ 2014, \aap, 564, A52

\bibitem[Lee et al.(2015)]{lee2015} Lee, N., Sanders, D.~B., Casey, C.~M., et al.\ 2015, \apj, 801, 80

\bibitem[Leroy et al.(2013)]{leroy2013} Leroy, A.~K, Walter, F., Sandstrom, et al.\ 2013, \apj, 146, 19




\bibitem[McGaugh(1991)]{mcgaugh1991} McGaugh, S.~S.\ 1991, \apj, 380, 140



\bibitem[Meynet et al.(2015)]{meynet2015} Meynet, G., Kudritzki, R.-P., \& Georgy, C.\ 2015, \aap, 581, A36


 


\bibitem[Pietrzy\'nski et al.(2006)]{pietrzynski2006} Pietrzy\'nski, G., Gieren, W., Soszy\'nski et al.\ 2006, \aj, 132, 2556

\bibitem[Pilyugin et al.(2014)]{pilyugin2014} Pilyugin, L.~S, Grebel, E.~K., Kniazev, A.~Y.\ 2014, \aj, 147, 132 

\bibitem[Przybilla et al.(2006)]{przybilla2006} Przybilla, N., Butler, K., Becker, S.~R., \& Kudritzki, R.~P.\ 2006, \aap, 445, 1099 

\bibitem[Puche et al.(1991)]{puche1991} Puche, D., Carignan, C., Wainscoat, R.~J.\ 1991, \aj, 101, 447




\bibitem[Robinson \& van Damme(1964)]{robinson1964} Robinson, B.~J., van Damme, K.~J.\ 1964, IAU Symposium Vol. 20, p. 287


\bibitem[Stasinska et al.(1986)]{stasinska1986} Stasinska, G., Comte, G., \& Vigroux, L.\ 1986, \aap, 154, 352

\bibitem[Stasinska et al.(2013)]{stasinska2013} Stasinska, G., Pena, M., Bresolin, F., \& Tsamis, Y.~G.\ 2013, \aap, 552, 12

\bibitem[Toribio San Cipriano et al.(2016)]{toribio2016} Toribio San Cpriano, L., Garcia-Rojas, J., Esteban, C. et al.\ 2016, \mnras, 458, 1866


\bibitem[T\"ullmann et al.(2003)]{tuellmann2003} T\"ullmann, R., Rosa, M.~R., Elwert, T. et al.\ 2003, \aap, 412, 69

\bibitem[Tremonti et al.(2004)]{tremonti2004} Tremonti, C.~A., Heckman, T.~M., Kauffmann, G., et al.\ 2004, \apj, 613, 898



\bibitem[Tully et al.(2009)]{tully2009} Tully, R.~B., Rizzi, L., Shaya, E.~J. et al.\ 2009, \aj, 138, 323

\bibitem[U et al.(2009)]{u2009} U, V., Urbaneja, M.~A., Kudritzki, R.-P., Jacobs, B.~A., Bresolin, F., \$ Przybilla, N.\ 2009, \apj, 704, 1120

\bibitem[Urbaneja et al.(2003)]{urbaneja2003} Urbaneja, M.~A., 
Herrero, A., Bresolin, F., et al.\ 2003, \apjl, 584, L73 

\bibitem[Urbaneja et al.(2005)]{urbaneja2005} Urbaneja, M.~A., 
Herrero, A., Kudritzki, R.-P., et al.\ 2005, \apj, 635, 311 

\bibitem[Urbaneja et al.(2008)]{urbaneja2008} Urbaneja, M.~A., 
Kudritzki, R.-P., Bresolin, F., et al.,\ 2008, \apj, 684, 118 

\bibitem[Urbaneja et al.(2011)]{urbaneja2011} Urbaneja, M.~A., 
Herrero, A., Lennon, D.~J., Corral, L.~J., \& Meynet, G.\ 2011, \apj, 735, 39 

\bibitem[Urbaneja et al.(2016)]{urbaneja2016} Urbaneja, M.~A., Kudritzki, R.~P., Gieren, W. et al.\ 2016, \apj, submitted

\bibitem[van de Steene et al.(2006)]{vansteene2006} van de Steene, G.~C., Jacoby, G.~H., Praet, C. et al.\ 2006, \aap, 455, 891

\bibitem[Vlajic et al.(2009)]{vlajic2009} Vlaji\'c, M., Bland-Hawthorn, J. \& Freeman, K.C. 2009, \apj, 697, 361


\bibitem[Webster \& Smith(1983)]{webster1983} Webster, B.~L., Smith, M.~G.\ 1983, \mnras, 204, 743 

\bibitem[Westmeier et al.(2011)]{westmeier2011} Westmeier, T., Braun, R., Koribalski, B.~S.\ 2011, \mnras, 410, 2217

\bibitem[Westmeier et al.(2013)]{westmeier2013} Westmeier, T., Koribalski, B.~S., Braun, R.\ 2013, \mnras, 434, 3511

\bibitem[Zahid et al.(2014)]{zahid2014} Zahid, H.~J., Dima, G.~I., Kudritzki, R.~P. et al.\ 2014, \apj, 791, 130

\bibitem[Zaritzky et al.(1994)]{zaritsky1994} Zaritzky, d., Kennicutt, R.~C., \& Huchra, J.~P.\ 1994, \apj, 420,87

\end{thebibliography}
\end{document}